# "Your click decides your fate": Leveraging clickstream patterns from MOOC videos to infer students' information processing & attrition behavior

A PROJECT REPORT

submitted by

**Tanmay Sinha (10BCE1114)**

*in partial fulfillment for the award*

of

**Bachelor of Technology**

degree in

**Computer Science and Engineering**

**School of Computing Science and Engineering**

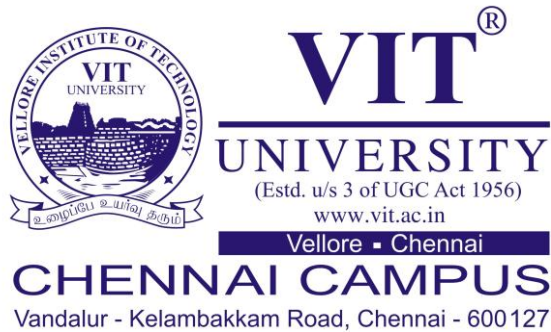

**July - 2014**

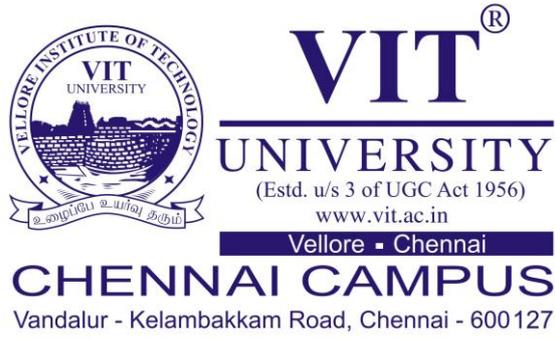

# School of Computing Science and Engineering

## DECLARATION

I hereby declare that the project entitled **"Your click decides your fate": Leveraging clickstream patterns from MOOC videos to infer students' information processing & attrition behavior** submitted by me to the School of Computing Science and Engineering, VIT University, Chennai Campus, Chennai 600127 in partial fulfillment of the requirements for the award of the degree of **Bachelor of Technology in Computer Science and Engineering** is a record of bonafide work carried out by me under the supervision of **Dr. Patrick Jermann (Executive Director, Center for Digital Education (CEDE), École Polytechnique Fédérale de Lausanne, Switzerland) and Dr. Pierre Dillenbourg (Professor, Computer Human Interaction Lab for Learning & Instruction (CHILI), École Polytechnique Fédérale de Lausanne, Switzerland).** I further declare that the work reported in this project has not been submitted and will not be submitted, either in part or in full, for the award of any other degree or diploma of this institute or of any other institute or university.

<div style="text-align: right;">
Signature<br>
**Tanmay Sinha (10BCE1114)**
</div>

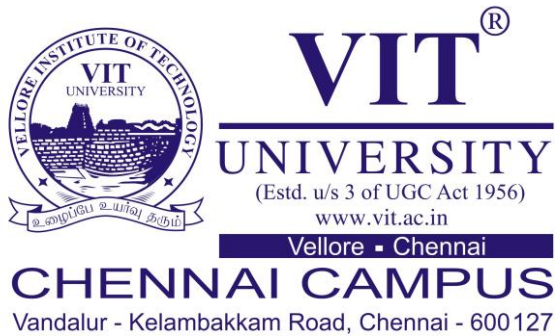

# School of Computing Science and Engineering

# CERTIFICATE

The project report entitled **"Your click decides your fate": Leveraging clickstream patterns from MOOC videos to infer students' information processing & attrition behavior** is prepared and submitted by **Tanmay Sinha (Register No: 10BCE1114).** It has been found satisfactory in terms of scope, quality and presentation as partial fulfillment of the requirements for the award of the degree of **Bachelor of Technology in Computer Science and Engineering** in VIT University, Chennai Campus, Chennai, India.

Signature    Signature
**Dr Patrick Jermann**    **Dr Indra Rajasingh**
Research Advisor, EPFL    Internal Guide, VIT

**Examined by**:

**Internal Examiner**    **External Examiner**

# ACKNOWLEDGEMENT

I would like to sincerely thank Dr. Patrick Jermann and Dr. Pierre Dillenbourg, for having given me the most wonderful opportunity to work towards my undergraduate thesis at École Polytechnique Fédérale de Lausanne (EPFL), Switzerland, one of the finest schools for Computer Science. Next, I would like to thank Dr Carolyn Rosé (Assistant Professor, Language Technologies Institute, Carnegie Mellon University(CMU)) for having given me a chance to initiate the exciting research on Massive Open Online Course in Summer 2013, and for helping to shape my interest towards computer supported collaborative learning. The exposure immensely helped me to draw diverse problem solving perspectives for my undergraduate thesis work. I want to acknowledge the support of lab colleagues, especially Nan Li (PhD candidate, EPFL) and David Adamson (PhD candidate, CMU) for their reflective, thought provoking discussions and brainstorming sessions.

Next, I would like to thank Dr. Indra Rajasingh (Internal guide & Professor, SAS) and Dr L Jegannathan (Dean, SCSE, VIT Chennai), who have been extremely supportive throughout my stay in VIT. No amount of appreciation for their guidance and understanding suffices. I would also like to express my gratitude towards professors at VIT, especially Dr S Hariharan, Dr Muthunagai, Professor Hannah Grace (SAS), Dr A Nayeemulla Khan and Professor K Vijaya (SCSE) for the love and care they have showered on me throughout my stay at VIT. I feel really lucky and proud to have the experience of interacting with such humble faculty in VIT.

At the end, I want to acknowledge the support of my parents, grandmother and my lovely sister, without whose encouragement and strength, I could not have reached where I am now.

# CONTENTS

# LIST OF TABLES

# LIST OF FIGURES

# LIST OF ABBREVIATIONS

| Abbreviation | Expansion |
|---|---|
| MOOCs | Massive Open Online Courses |
| IPI | Information Processing Index |
| VWSS | Video Watching State Sequence |

# ABSTRACT

With an expansive and ubiquitously available gold mine of educational data, Massive Open Online courses (MOOCs) have become the an important foci of learning analytics research. The hope is that this new surge of development will bring the vision of equitable access to lifelong learning opportunities within practical reach. MOOCs offer many valuable learning experiences to students, from video lectures, readings, assignments and exams, to opportunities to connect and collaborate with others through threaded discussion forums and other Web 2.0 technologies. Nevertheless, despite all this potential, MOOCs have so far failed to produce evidence that this potential is being realized in the current instantiation of MOOCs. In this work, we primarily explore video lecture interaction in Massive Open Online Courses (MOOCs), which is central to student learning experience on these educational platforms. As a research contribution, we operationalize video lecture clickstreams of students into behavioral actions, and construct a quantitative information processing index, that can aid instructors to better understand MOOC hurdles and reason about unsatisfactory learning outcomes. Our results illuminate the effectiveness of developing such a metric inspired by cognitive psychology, towards answering critical questions regarding students' engagement, their future click interactions and participation trajectories that lead to in-video dropouts. We leverage recurring click behaviors to differentiate distinct video watching profiles for students in MOOCs. Additionally, we discuss about prediction of complete course dropouts, incorporating diverse perspectives from statistics and machine learning, to offer a more nuanced view into how the second generation of MOOCs be benefited, if course instructors were to better comprehend factors that lead to student attrition. Implications for research and practice are discussed.

# CHAPTER 1

# INTRODUCTION

## 1.1 Massive Open Online Courses(MOOCs)

Mushrooming as a scalable lifelong learning paradigm, Massive Open Online Courses (MOOCs) have enjoyed significant limelight in recent years, both in industry and academia (Haggard et al., 2013). The rationale behind the design of these MOOCs is the underlying theory of connectivism (Kop and Adrian, 2008), which stresses more on interaction with other participants and on the student-information relationship. As MOOCs continue to proliferate in the realm of online education, we expect such a form of lifelong learning holding tremendous potential, to provide students with cognitive surplus beyond traditional forms of tutelage. The euphoria is about the transformative potential of MOOCs to revolutionize online education (North et al., 2014), by connecting and fostering interaction among millions of learners who otherwise would never have met, and providing autonomy to these learners to grapple with the course instruction at their own pace of understanding.

However, despite this expediency, there is also considerable skepticism in the learning analytics research community about MOOC productiveness (Nawrot and Antoine, 2014), primarily because of unsatisfactory learning outcomes that plague these educational platforms and induce a funnel of participation (Clow, 2013). Because of the option of free and open registration, a publicly shared curriculum and open ended outcomes, very often it has been observed that there is massive enrollment in these digitalized MOOC courses. However, participation in a MOOC is "emergent, fragmented, diffuse, and diverse" (McAulay et al., 2010). The extremely high rates of attrition that have been reported for this first generation of MOOCs, is of great concern.

## 1.2 Motivation

With a "one size fits all" approach that MOOCs follow, scaled up class sizes, and lack of face to face interaction coupled with such high student teacher ratios (Guo and Katharina, 2014), students' motivation to follow the course oscillates (Davis et al., 2014). This is comprehensibly reflected in escalating attrition rates in MOOCs, ever since they have started maturing (Belanger and Jessica, 2013; Schmidt and Zach, 2013; Yang et al., 2013). Supporting the participation of these struggling students may be the first low hanging fruit for increasing the success rate of courses. Because it is not feasible for MOOC instructors to manually provide individualized attention that caters to different backgrounds, diverse skill levels, learning goals and preferences of students, there is an increasing need make directed efforts towards automatically providing better personalized content in e-learning (Sinha et al., 2013; Lie et al., 2014; Sinha, 2014a). The provision of guidance with regard to the organization of the study and regulation of learning is a domain that also needs to be addressed.

A prerequisite for such an undertaking is that we, as MOOC researchers, understand how diverse ecologies of participation develop as students interact with the course material (Fischer, 2011), and how learners distribute their attention with multiple forms of computer mediated inputs in MOOCs. This would help to better their experience of participation along the way as they struggle and then ultimately drop out, for example by examining participation rates of collaborations through group mirrors and metacognitive tools that dynamically display students' progress & help in interaction regulation (Jermann and Dillenbourg, 2008).

While substantial research has been done on studying MOOC discussion forums (Ramesh et al., 2013; Brinton et al., 2013; Anderson et al., 2014; Sinha, 2012;Sinha, 2014b), grading strategies for assignments (Tillmann et al., 2013; Kul et al., 2014) and deployment of reputation systems for MOOCs (Coetzee et al., 2014), inner workings of students' interaction while watching MOOC video lectures have been much less focused upon. Current published analyses of participation in MOOCs have not provided the needed visibility into the interactions of students within the MOOC context.

Given that roughly 5% (Huang et al., 2014) of students actually participate in MOOC discussion forums, it would be legitimate to ask whether choosing video lectures as units of analysis would be more insightful. After 330,000 registrations in MOOC courses at EPFL in 2013, our experience reflects that out of the 100% students who register, 75% show up: 50% of them primarily watch video lectures and the rest 25% additionally work out homeworks and assignments. Thus, majority of students have video lecture viewing as their primary MOOC activity. Outlining the perspectives on studying learner interaction in MOOC, that have been explored so far in the emerging literature on MOOCs, we find that most of this work has grown out of research on distance education that preceded the emergence of MOOCs as an online learning paradigm.

**1.3 Our Current Research Overview**

Video lectures form a primary and an extremely crucial part of MOOC instruction design. They serve as gateways to draw students into the course. Concept discussions, demos and tutorials that are held within these short video lectures, not only guide learners to complete course assignments, but also encourage them to discuss the taught syllabus on MOOC discussion forums. Prior work has investigated how video production style (slides, code, classroom, khan academy style etc) relates to students' engagement (Guo et al., 2014a) and examined what features of the video lecture and instruction delivery, such as slide transitions (change in visual content), instructor changing topic (topic modeling and ngram analysis) or variations in instructor's acoustic stream (volume, pitch, speaking rate), lead to peaks in viewership activity (Kim et al., 2014).There has been increasing focus on analyzing raw click-level interactions resulting from student activities within individual MOOC videos (Guo et al., 2014b). However, to the best of our knowledge, we present the first study that describes usage of such detailed clickstream information to form cognitive video watching states that summarize student clickstream. Instead of using summative features that express student engagement, we leverage recurring click behaviors of students interacting with MOOC video lectures, to construct their video watching profile. To an extent, clickstreams simplify computing student retention, since a large variety of interactions could potentially indicate continued interest in a course.

Based on these richly logged interactions of students, we develop computational methods that answer critical questions such as a)how long will students grapple with the course material and what will their engagement trajectory look like, B)what future click interactions will characterize their behavior, C)whether students are ultimately going to survive through the end of the video. As an effort to improve the second generation of MOOC offerings, we perform a hierarchical three level clickstream analysis, deeply rooted in foundations of cognitive psychology.

Incidentally, we explore at a micro level whether, and how, cognitive mind states govern the formation and occurrence of micro level click patterns. Towards this end, we also develop a quantitative information processing index and monitor its variations among different student partitions that we define for the MOOC. Such an operationalization can help course instructors to reason how students' navigational style reflects cognitive resource allocation for meaning processing and retention of concepts taught in the MOOC. Furthermore, we delineate a methodology to group students and unveil distinct patterns of video lecture viewing.

**1.4 Study Context**

The data for our current study in this thesis comes from an introductory programming MOOC "Functional Programming in Scala" that was offered on the Coursera MOOC platform in 2012. This MOOC comprises of 48 video lectures (10 Gb of Json data), which has been parsed and preprocessed into a convenient format for experimentation. In these interaction logs, every click of users on the MOOC video player is registered (play, pause, seek forward, seek backward, scroll forward, scroll backward, ratechange). We have information about the rate at which video is played, total time spent on playing the video and time spent on/in-between various click events such as play, pause, seek etc. This data allows us to investigate how students watch the video, for example, where do they pause and for how long, which parts of the video were looked at more than one time and which parts were skipped etc

**1.5 Organization of this thesis**

In the remainder of this thesis, we motivate our three level hierarchical MOOC video clickstream analysis (operations, actions, information processing activities), describing related work along the way and technical approach followed in section 2. In Chapter 3, we perform our first set of validation experiments, by setting up certain machine learning experiments, specifically engagement prediction, next click state prediction and in-video dropout prediction. As a second set of experiments to validate our developed methodology, Chapter 4 provides details of how the developed information processing index varies among different student partitions in the MOOC, hinting at how significant are the differences. In Chapter 5, we talk about formation of an informal clickstream network, specifically discussing Markov clustering based and Social network based modeling approaches to coherently illuminate distinct cognitive video watching preferences of students. In Chapter 6, we discuss about prediction of Complete Course Dropouts to offer a more nuanced view into how the second generation of MOOCs be benefited, if course instructors were to better comprehend factors that lead to student attrition. Implications for future work and conclusion is presented in Chapter 7.

# CHAPTER 2

# OPERATIONALIZING THE CLICKSTREAM

## 2.1 Level 1 (Operations)

From our raw clickstream data, we construct a detailed encoding of students' clicks in the following 8 categories: Play (Pl), Pause (Pa), SeekFw (Sf), SeekBw (Sb), ScrollFw (SSf), ScrollBw (SSb), RatechangeFast (Rf), RatechangeSlow (Rs). When two seeks happen within a small time range ($< 1$ sec), we group these seek events into a scroll. Additionally, to encode 'Rf' and 'Rs', we look for the playrate of the click event that occurs just before the 'Ratechange' click and compare it with students' currently changed playrate, to determine whether he has fastened/slowed down his playing speed. As a next step, we concatenate these click events for every student, for every video lecture watched. This string of symbols that characterizes the sequence of clickstream events is referred to as a 'video watching state sequence'. For e.g: PlPaSfSfPaSbPa.., PlSSbPaRsRsPl..

The reason behind encoding clickstreams to such specific categories, accommodating scrolling behavior and clicks representative of increase and decrease in video playing speed, is to experimentally analyze and understand the impact of such a granularity on our experiments, which are designed with an objective to capture the motley of differently motivated behavioral watching style in students.

## 2.2 Level 2 (Behavioral Actions)

Existing literature on web usage mining says that representing clicks using higher level categories/concepts, instead of raw clicks, better exposes the browsing pattern of users. This might be because high level categories have better noise tolerance than naive clickstream logs. The results obtained from grouping clickstream sequences at per click resolution are often difficult to interpret, as such a fine resolution leads to a wide variety of sequences, many of which are semantically equivalent. To tackle this problem and get more insights into student behavior in MOOCs, the clicks can be first grouped into categories based on suitable metadata information, and then the sequences can be formed from the concept category of the click events present in the sequences. Doing this would reduce the sequence length that would be more easily interpretable.

There is some existing literature (Banerjee and Ghosh, 2000; Wang et al., 2013), that just considers click as a binary event (yes/no) and discusses formation of concept based categories based on the area/sub area of the stimulus where the click was made. However, in our MOOC data, because of absence of metadata about the clicks, it would be more meaningful to form such behavioral categories from these click categories itself, which are encoded at very fine granularity.

Therefore, to summarize a students' clickstream, we obtain the n-grams with maximum frequency from the clickstream sequence (a contiguous sequence of 'n' click actions). Such a simple n-gram representation convincingly captures the most frequently occurring click actions that students make in conjunction with each other (n=4 was empirically determined as a good limit on clickstream subsequence over specificity). Then, we construct seven semantically meaningful behavioral categories using these n-grams, selecting representative click groups that occur within top 'k' most frequent n-grams (k=100). Each behavioral category acts like a latent variable, which is difficult to measure from data directly. We exclude the n-gram sequences having only 'play' and 'pause' click actions in the clickstream.

- **Rewatch:** PlPaSbPl, PlSbPaPl, PaSbPlSb, SbSbPaPl, SbPaPlPa, PaPlSbPa
- **Skipping:** SfSfSfSf, PaPlSfSf, PlSfSfSf, SfSfSfPa, SfSfPaPl, SfSfSfSSf, SfSfSSfSf, SfPaPlPa, PlPaPlSf
- **Fast Watching:** PaPlRfRf, RfPaPlPa, RfRfPaPl, RsPaPlRf, PlPaPlRf (click group of Ratechange fast clicks while playing or pausing video lecture content, indicating speeding up)

- **Slow Watching:** RsRsPaPl, RsPaPlPa, PaPlRsRs, PlPaPlRs, PaPlRsPa, PlRsPaPl (click group of Ratechange slow clicks while playing or pausing video lecture content, indicating slowing down)
- **Clear Concept:** PaSbPlSSb, SSbSbPaPl, PaPlSSbSb, PlSSbSbPa (a combination of SeekBw and ScrollBw clicks, indicating high tussle with the video lecture content)
- **Checkback Reference:** SbSbSbSb, PlSbSbSb, SbSbSbPa, SbSbSbSf, SfSbSbSb, SbPlSbSb, SSbSbSbSb (a wave of SeekBw clicks)
- **Playrate Transition:** RfRfRsRs, RfRfRfRs, RfRsRsRs, RsRsRsRf, RsRsRfRf, RfRfRfRf (a wave of ratechange clicks)

In an attempt to quantify the importance of each behavioral action in characterizing the clickstream, we adopt a fuzzy string matching approach. The advantage with such an approach over simple "vector of proportions" representation, is that a weight (based on similarity of click groups present in each behavioral category, with the full clickstream sequence) is assigned to each of the grouped behavioral patterns for a given students' video watching state sequence. The fuzzy string method (Van, 2014) is justified because it caters to the noise that might be present in raw clickstream logs of students, in six different ways, as mentioned in Table 1. After identifying these cases and meticulous experimental evaluation, we apply the following distance metrics and tuning parameters: Cosine similarity metric (1- Cosine distance; figure 1) between the vector of counts of n-gram (n=4) occurrences for Cases 1 and 2, Levenshtein similarity metric (1- Levenshtein distance; figure 2) for Cases 3 (weight for deletion=0, weight for insertion and substitution=1), 4, 5, 6 (weight for deletion=0.1, weight for insertion, substitution=1) capture all these six intuitions.

$$d_{\cos}(s, t; q) = 1 - \frac{v(s; q) \cdot v(t; q)}{\|v(s; q)\|_2 \|v(t; q)\|_2}$$

Fig 1: Q-gram based cosine distance measure. $v(s; q)$ is a nonnegative integer vector whose coefficients represent the number of occurrences of every possible q-gram in s.

$$d_{lv}(s,t) = \begin{cases} 0 \text{ if } s = t = \varepsilon \\ \min\{ \\ \quad d_{lv}(s, t_{1:|t|-1}) + w_1, \\ \quad d_{lv}(s_{1:|s|-1}, t) + w_2, \\ \quad d_{lv}(s_{1:|s|-1}, t_{1:|t|-1}) + [1 - \delta(s_{|s|}, t_{|t|})]w_3 \\ \} \text{ otherwise.} \end{cases}$$

Fig 2: Levenshtein distance measure. w1, w2 and w3 are the nonnegative penalties for deletion, insertion, and substitution when turning t into s.

As a next step, all subcategories of click groups that lie within each behavioral category, are aggregated by summing up the individual fuzzy string similarity weights. Then, we perform a discretization of these summed up weights, for each behavioral category, by equal frequency (High/Low). The concern of adding up two distance metrics that do not lie in the same range, is thus alleviated, because the dichotomization automatically places highly negative values in the "Low" category and positive values closer to 0 in the "High" category. This results in a clickstream vector, where every element of the vector tells us about the weight (importance) of a behavioral category for characterizing the clickstream. Thus, the output from Level 2 is such a summarized clickstream vector. For e.g: (Skipping=High, Fast Watching=High, Checkback Reference=Low, Rewatch=Low, ....).

### 2.3 Level 3 (Information Processing)

Watching MOOC videos is an interaction between the student and the medium, and therefore the conceptualization of higher-order thinking eventually leading to knowledge acquisition (Chi, 2000), is under control of both the student (who decides what video segment to watch, when/in what order to watch, how hard an effort be made to try and understand a specific video segment) and the medium/video lecture (the content/features of which decides what capacity allocation is required by the student to fully process the information contained).

| Condition | Case | Clickstream A | Clickstream B | Fuzzy string matching verdict |
|---|---|---|---|---|
| **Full string match** | 1: Varying clickstream Length | PlPaPlSfPaSfSbSbPl | PlPaPlSfPaSfSbSbPlPaSbSbRfRs | Weight(P,A)>Weight(P,B) |
| | 2: Behavioral pattern appears more than once | PlPaPlSfPaSfSbSbPl | PlPaPlSfPaSfSbSbPlPlSfPaSf | Weight(P,A)< Weight(P,B) |
| **No match** | 3: No appearance of behavioral pattern | RfSbSbRs | SSfSSfRsSfSfSfRfRfRfRfRf | Weight(P,A) ≠ Weight(P,B) |
| **Partial string match** | 4: Variation in number of individual clicks | RfSbSbRsPlSbPaSb | RfSbSbRsPlSbSfPaSfSb | Weight(P,A)<Weight(P,B) |
| | 5: Variation in scattering of individual clicks | RfSbSbRsPlSbSfPaSfSb | RsPlSbSSbSfPlSbRsRsPaSbRfSf | Weight(P,A)>Weight(P,B) |
| | 6: Reverse order of individual click appearance | RfSbSbRsSbSfPaSfSbPl | RfSbSbRsPlSbSfPaSfSb | Weight(P,A)<Weight(P,B) |

Table 1: Fuzzy string similarity weights for the sample behavioral action P("PlSfPaSf")
Weight(P, A/B) represents the similarity of the pattern P w.r.t string A/B.

Research has consistently found that the level of cognitive engagement is an important aspect of student participation (Carini et al., 2006). The cognitive processing is influenced by the appetitive (approach) and aversive (avoidance) motivational systems of a student, which activate in response to motivationally relevant stimuli in the environment (Cacioppo and Gardner, 1999). In the context of MOOCs, the appetitive system's goal is in-depth exploration and information intake, while the aversive system primarily serves as a motivator for not attending to certain MOOC video segments. Thus, click behaviors representative of appetitive motivational system are rewatch/clear concept/slow watching, while click behaviors representative of aversive motivational system are skipping/fast watching.

In this work, we try to construct students' information processing index, based on the "**Limited Capacity Information Processing Approach**" (Basil, 1994; Lang et al., 1996; Lang, 2000), which asserts that people independently allocate limited amount of cognitive resources to tasks from a shared pool.

Before explaining the dynamic process of human cognition through the lens of this model, we must be aware of the following two assumptions: A)People are limited capacity information processors. In case of cognitive overload, processing suffers, B)The sub-processes involved in information processing pipeline occur constantly, continuously and simultaneously.

When students build a mental representation of the information presented in a MOOC video lecture segment, it is not precise. To what extent different subprocesses in the pipeline share information, or, which subprocesses make the largest resource demands, depends on students' prior knowledge/skill level, motivations for joining the course and outcomes sought. Because students choose bits of information (specific content) to process and encode, therefore they navigate the videos in non linear fashion.

Moreover, students in MOOCs can adjust the speed of information processing (by pausing, seeking forward/backward, ratechange clicks). Therefore, time sensitive subprocesses in the pipeline (depicted in figure 3) seem compatible with this notion. Video watching in MOOCs requires students to recall facts that they already know, so as to follow and comprehend the concept being currently taught. So, depending on the a)expertise level, which decides how available the past knowledge is and how hard is it to retrieve the previously known facts, b)perception of video lecture as difficult or simple to understand, c)motivation to learn or just have a look at the video lecture, cognitive resource allocation would vary among these various subprocesses. This in turn, would be reflected by the underlying nature of clicks students make, which serve as responses to the stimuli.

Consider an example of students who watch the MOOC lecture, primarily because of reasons such as gaining familiarity with the topic. Such students would purposely not allocate their processing resources to "memory" part of the information processing pipeline (encode, store, retrieve). Additionally, they will decode and process minimal information that is required to follow the story. On the contrary, students who watch the MOOC lecture, with the aim of scoring well in post-tests (MOOC quizzes and assignments), would allocate high cognitive processing to understand, learn and retain information from the lecture. Thus, such students would process information more fully and thoroughly, despite a possibility of cognitive overload. In order to relate our behavioral actions constructed from the raw clickstream with this rich and informative stream of literature, we buttress our "Information Processing Index (IPI)" development, on the following arguments. Figure 3 summarizes the clarifications described below:

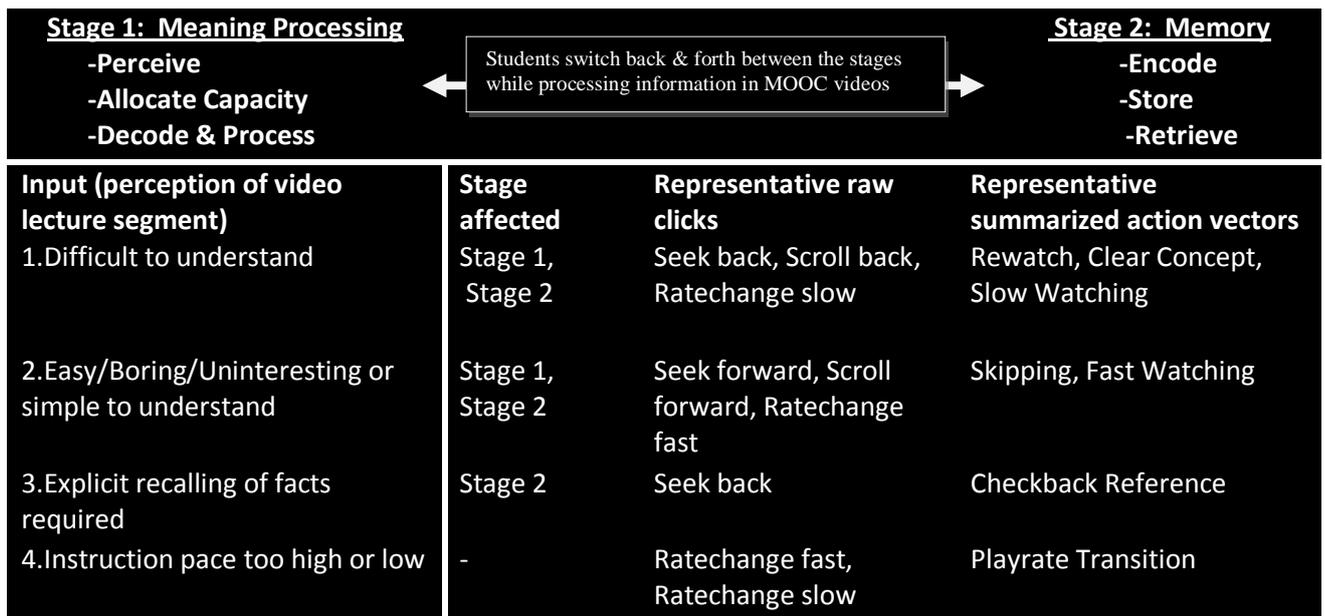

Fig 3: Relating students' information processing to click behaviors exhibited in the MOOC, based on video lecture perception

- When students perceive certain video lecture segment as difficult, they allocate more capacity (cognitive resources) to repeatedly decode, process and store information. Students such as these who **"rewatch"** or try to **"clear their concept"** are more likely go through the pipeline stages sequentially, rather than simultaneously (high information processing involved)
- When students perceive certain video lecture segment as easy/boring/uninteresting, they allocate very minimal/no capacity to process, decode and store information in memory. Such **"skipping"** behavior involves low information processing.

- When students perceive certain video lecture segment as simple to understand (perhaps because they are already familiar with concept being taught), they allocate comparatively lesser capacity than normal/regular watching, and comparatively more capacity than completely skipping video segment to process, decode, encode and store the video segment information. Such students who exhibit **"fast watching"** in their clickstream are likely to do low information processing overall.
- When students perceive certain video lecture segment as difficult to understand (perhaps because some tough concept is being taught), they need to allocate comparatively higher capacity (processing resources) than normal watching to process, decode, encode and store the video segment information. Such students who exhibit **"slow watching"** in their clickstream are likely to do high information processing overall.
- Students might **check back for reference** in the following two cases in MOOCs. For both these cases, "meaning processing" (Stage 1) part of the pipeline is likely to be processed normally. Thus, the problem is more probable to occur in "memory" (Stage 2) part of the pipeline (i.e., not in the information processing, but the outcome of the information processing). So, cognitive resource allocation should be comparatively higher than skipping/fast watching (because Stage 1 of information processing has been successfully done), but because information processing is still low in Stage 2, this action should be weighted negative overall. Such students who exhibit "checking back for reference" in their clickstream are likely to do low information processing.
    - A) If a previously taught concept is referred, and student had not paid sufficient attention previously, but is aware of such a concept being mentioned earlier, he has to refer back

- (problem in encoding/recognition stage, less resources allocated to this step, therefore poor memory for detail)
  - o B)If a previously taught concept is referred which happens to be, for example, some complex formulae, it is not expected of a student to exactly remember the formulae. Therefore even though he might have paid high attention to encode the information earlier, storage would have been shorted at that time (shared resource pool). Therefore, the student might not be able to concurrently retrieve the information now and has to refer back (problem in storage and retrieval stages, less resources allocated to these steps, therefore information poorly stored) (Lang and Basil, 1998)
- Students can adjust and get to their comfort level of video watching speed while watching video lectures in MOOCs. Though some amount of cognitive processing is involved to determine the pace at which the MOOC instruction and students' understanding will be coherent, this group of click behavior is not directly related to the actual processing of information content. A **"playrate transition"** just determines the speed at which a student wants to process information. So, such a behavioral action could be considered neutral.

In order to relate our behavioral actions constructed from the raw clickstream with this rich and informative stream of literature, we create a taxonomy of behavioral actions exhibited in the clickstream to construct a quantitative "Information Processing Index (IPI)".The above established hierarchy of information processing is summarized in Figure 4. Negative weights are necessary to distinguish between the "high" and "low" weights for each behavioral action. For example, if skipping=high is weighted -3, skipping=low will be weighted +3 on the information processing index. Using these linear weight assignments, we define students' information processing index as follows:

**Information Processing Index (IPI)** $= (-1)^j \sum_{i=1}^{7}$ WeightAssign (Behavioral Action i), j=1,2 depending on whether the behavioral action is weighted low or high.

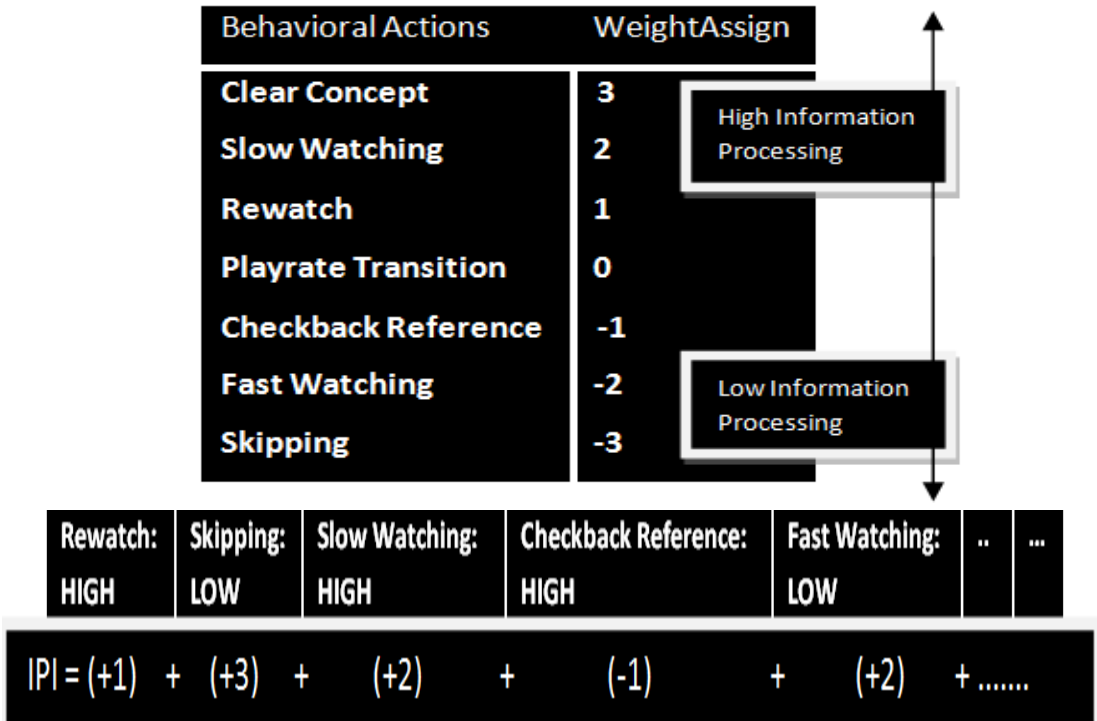

Fig 4: Linear weight assignments for behavioral clickstream actions, according to the information processing hierarchy developed

One of the focal utilities of developing such a quantitative index is that meaningful intervention could be provided in real time to students, as they steadily build up their video watching profile while interacting with MOOC video lectures. When IPI > 0, it can be inferred that high information processing is being done by students. Therefore MOOC instructors need to check for coherency in pace of instruction delivery and students' understanding. This might also hint towards redesigning specific video lecture segments and simplifying them so that they become easier to follow. On the contrary, when IPI < 0, low information processing is being done by students. Therefore MOOC instructors need to help students better engage with the course, by providing them additional interesting reading/assignment material, or fixing video lecture content such that it captures students' attention. The neutral case of IPI = 0 occurs when students' locally exhibited high and low information processing needs in their evolving clickstream sequence counterbalance each other. So, interventions need to made depending on the video lecture segment, where IPI was >0 or <0.

# CHAPTER 3

# VALIDATION EXPERIMENTS 1: MACHINE LEARNING

We use machine learning to validate the methodology developed in Section 2.1 and 2.2 for summarizing students' clickstream. The motivation behind setting up these experiments is to automatically measure students' length of interaction with MOOC video lectures, understand how they develop their video watching profile and discern what viewing profile of students leads to in-video dropouts. Furthermore, we validate the methodology developed in section 3.3 by statistically analyzing variations of IPI and testing its sensitivity to student attrition using survival models.

## 3.1 Preliminaries on Machine Learning

Machine learning, a branch of artificial intelligence, concerns the construction and study of systems that can learn from data. The core of machine learning deals with representation and generalization. Representation of data instances and functions evaluated on these instances are part of all machine learning systems. Generalization is the property that the system will perform well on unseen data instances; the conditions under which this can be guaranteed are a key object of study in the subfield of computational learning theory. A computer program is said to learn from experience E with respect to some class of tasks T and performance measure P, if its performance at tasks in T, as measured by P, improves with experience E.

Supervised learning is the machine learning task of inferring a function from labeled training data. The training data consist of a set of training examples. In supervised learning, each example is a pair consisting of an input object (typically a vector) and a desired output value (also called the supervisory signal). A supervised learning algorithm analyzes the training data and produces an inferred function, which can be used for mapping new examples. An optimal scenario will allow for the algorithm to correctly determine the class labels for unseen instances.

One of the popular supervised learning algorithms that has been found to work well with text data is Logistic Regression (figure 5). It is a discriminitive and probabilistic classification model. By discriminative, we mean that the algorithm assumes some functional form for P(Y|X) or for the decision boundary, and estimates parameters of P(Y|X) directly from training data. This is unlike the generative Naive Bayes model, that assumes some functional form for P(X|Y) and P(Y), estimates parameters of P(X|Y), P(Y) directly from training data and uses Bayes rule to calculate P(Y|X). High coefficient weights in Logistic Regression may lead to overfitting. So, we also apply L2 regularization in our approach. Regularization works by adding penalty associated with high coefficient values. L1 usually corresponds to setting a Laplacean prior on the regression coefficients and picking a maximum a posteriori hypothesis. L2 similarly corresponds to Gaussian prior. L2 regularization is expected to do better for our case because it is directly related to minimizing the VC dimension of the learned classifier (capacity/complexity of a classifier: no. of pts that can be shattered)

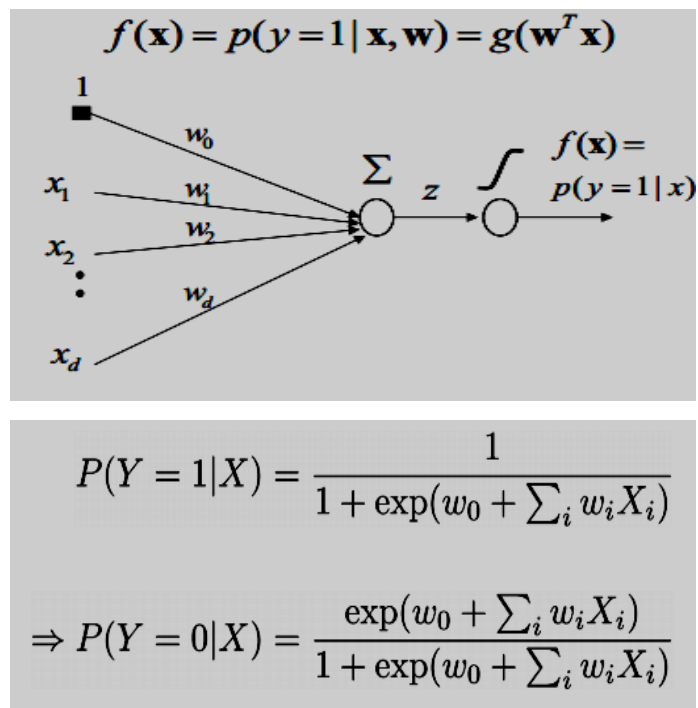

Fig 5: Summary of the Logistic Regression model

To evaluate the performance of machine learning algorithms, we use the following 3 metrics: A)Classification Accuracy, B)Kappa, C) False Negative Rate(FNR). Ideally, we want A and B to be high, and C to be low. These metrics are explained with an example in table 2 below:

| Actual/Predicted | Positive Class | Negative Class | Total |
|---|---|---|---|
| Positive Class | 60 | 20 | 80 |
| Negative Class | 40 | 25 | 65 |
| Total | 100 | 45 | |

Table 2: An example to explain Machine Learning Performance metrics

- **Accuracy** = $P_0$ = (60+25)/(60+20+40+25) = 0.586

    { Correctly classified examples}

- **Kappa** = $(P_0 - P_e^C)/(1 - P_e^C)$

    $P_e^C$ = (100/145)* (80/145) + (45/145)*(65/145) = 0.517

    So, Kappa = 0.586-0.517/1-0.517=0.142

    {$P_0$ represents the probability of overall agreement over the label assignments between the classifier and the true process, and $P_e^C$ represents the chance agreement over the labels and is defined as the sum of the proportion of examples assigned to a class times the proportion of true labels of that class in the data set}

- **False Negative Rate(FNR)** = 40/(40+25) = 0.615

    {Instances which are falsely predicted as negative}

### 3.2 Machine Learning Experiment Design

Students, while watching MOOC video lectures can pause, seek, scroll and change the rate of the video. Thus, it is meaningful to quantify students' engagement as the summation of video playing time, seeks & pauses multiplied by the playback rate. For example, if a student plays 700 secs out of a 1000 sec video, pauses 2 times for 100 secs each, at an average play rate of 1.5, he effectively engages with the video for (700+200)*1.5=1350 secs. Such an interaction measure multiplied by playback rate, is representative of effective video lecture content covered.

**Research Question 1 (How much do you engage?)**: Can students' clickstream sequence be used to predict how long will students be interacting with the video lecture?

**Settings**: The data for this experiment comes from a randomly chosen video lecture 4-6 (6th lecture in the 4th week of the course with not too many initial lurkers and not too many dropouts). For experimental purposes, engagement times for students are discretized by equal frequency into 2 categories (High/Low). The dependent variable is student engagement (High: 1742 examples, Low: 1741 examples). L2 regularized Logistic Regression is used as the training algorithm (with 10 fold cross validation annotated by student-id and rare feature extraction threshold being 2). As features, we extract Ngrams of length 4 and 5, string length and regular expressions from students' clickstream sequences. In the changed setup, we consider summarized behavioral category vectors (output from level 2) as column features.

Next, we see focus our attention on how clickstream sequences evolve. If we know that a students' interaction with the video lecture is going to be for a long time (reflected by high engagement), it might have been the case that he was struggling at the current level of instruction (for example, a high combination of pause/seek backward events). Therefore, if this phenomenon can be detected in real time video lecture interaction, such learners can be presented with reinforcement course material before moving forward. Alternatively, if we know that a students' interaction with the video lecture is going to be for a short time (reflected by low engagement), the student might be bored or is quite likely to skip course content forward often. Such students could be presented with advanced study material. However, in order to develop such a real time knowledge model and tailor targeted interventions at the student, we need to study the trajectory of click sequence formation.

**Research Question 2 (Are you bored or challenged?)**: Can we precisely predict what will be the next sequence of clicks that leads students to different engagement states?

**Settings**: The data for this experiment comes from video lecture 4-6 (6th lecture in the 4th week of the course). The dependent variable is the next click state of students (Pa, Pl, Sf, SSf, Sb, SSb, Rf, Rs). L2 regularized Logistic Regression is used as the training algorithm (with 5 fold cross validation annotated by student-id and rare feature extraction threshold being 5). If we want to predict the click at the $i^{th}$ instant, we extract the following features from 0 till $(i-1)^{th}$ instant: A)Engagement with the video lecture as defined for Research Question 1(High/Low); B)Proportion of click events belonging to Pl/Pa/Sf/SSf/Sb/SSb/Rf/Rs (representative of kind of interaction with the stimulus); C)N-grams of length 4,5 and string length from students' clickstream sequences. In the changed setup, we consider summarized behavioral category vectors (output from level 2) as column features.

As students progress through the video, they slowly build up their video watching profile, by interacting with the stimulus in different proportions, which in turn depend on their click action sequences. This motivates our next machine learning experiment, which seeks to derive utility from the first two experiments. Navigating away from the video without completing it fully is an outcome of low student engagement. A student is more likely to watch till the end of a video lecture, if the presentation activates his thinking. Thus, it would be interesting to see, whether the nature of students' interaction provide us a hint about such in-video dropouts. Prior work has made a preliminary study on how in-video dropout is correlated with length of the video, and how in-video dropout varies among first time watchers and rewatchers (Guo et al., 2014b). However, we consider video interaction features at a much finer granularity, representative of how students progress through the video. In doing so, we use detailed clickstream information, including 'Seekfw', 'Seekbw' and 'RateChange' behavior, in addition to merely play/pause information.

**Research Question 3 (Will you drop out of the video?)**: What video watching profile of students leads to in-video dropouts?

**Settings**: The data for this experiment comes from video lecture 4-6 (6th lecture in the 4th week of the course). The dependent variable is the binary variable, in-video dropout (0/1). To address the skewed class distribution, cost sensitive L2 regularized Logistic Regression is used as the training algorithm (with 10 fold cross validation annotated by student-id and rare feature extraction threshold being 2). To extract the interaction footprint of a student before he drops out of the video, we extract the following features: A)N-grams of length 4,5 and string length from students' click stream sequences; B)Proportion of click events belonging to Pl/Pa/Sf/SSf/Sb/SSb/Rf/Rs (representative of kind of interaction with the stimulus); C)Engagement with the video lecture as defined for Research Question 1(High/Low); D)Last click action before dropout happened; E)Time spent after the last click action was made (discretized by equal frequency to High/Low). In the changed setup, we consider summarized behavioral category vectors (output from level 2) as column features.

### 3.3 Results

Results of the three machine learning experiments, along with the most representative (weighted) features that characterize classes, are reported in table 3. False negative rate is lower for Case 1.B and Case 3.B, as compared to Case 1.A and Case 3.A, which shows the effectiveness of the clickstream summarization approach in pre-deciphering the fate of students to some extent.

| Research Question | Condition | Accuracy Kappa | False Negative Rate | Most representative (weighted) features that characterize classes |
|---|---|---|---|---|
| **Engagement Prediction** | 1) Raw Clicks | **0.81** **0.63** | **0.24** | **High** (skipping=low, playrate transition=low, rewatch=high, slow watching=low, checkback reference=low, clear concept=high) |
| | 2) Summarized behavioral action vectors | **0.75** **0.49** | **0.15** | **Low** (skipping=high, playratetransition=high, rewatch=low, slow watching=high, checkback reference=high, clear concept=low) |
| **Next Click Prediction** | 1) Raw Clicks | **0.68** **0.57** | - | **SeekFw** (playratetransition=low, skipping=low, fast watching=high, clearconcept=low) |
| | 2) Summarized behavioral action vectors | **0.66** **0.54** | - | **SeekBw** (checkbackreference=high, rewatch=low, playratetransition=low, propSeekBw, clearconcept=high) |
| | | | | **Ratechangefast** (playratetransition=high, rewatch=low, checkbackreference=low) |
| | | | | **Ratechangeslow** (playratetransition=high, clearconcept=high) |
| **In-Video dropout Prediction** | 1) Raw Clicks | **0.90** **0.69** | **0.19** | **Non dropouts** (skipping=low, clearconcept=high, slow watching=high, Checkbackreference=low, rewatch=high, engagementfromStart=low, engagementlastClick=high) |
| | 2) Summarized behavioral action vectors | **0.90** **0.70** | **0.15** | **Dropouts** (skipping=high, clearconcept=low, slowwatching=low, engagementfromStart=high, rewatch=low, engagementlastClick=low, checkbackreference=high) |

Table 3: Performance metrics for our machine learning experiments. Random baseline performance is 0.5 (50%)

# CHAPTER 4

# VALIDATION EXPERIMENTS 2: IPI VARIATIONS

To see how IPI fluctuates among different student partitions, and validate whether our operationalization produces meaningful results, we do extensive statistical analysis, specifically z tests (to test significance of difference between mean of 2 samples drawn from same population{population standard deviation known}), computing one way anova measures (to test significance of difference between more than 2 sample means) and performing posthoc tests.

$|Z| = [\overline{x1} - \overline{x2}] \div [\sigma * sqrt\{(1 \div n2) + (1 \div n1)\}]$, where $\overline{x1}$ is the mean of sample 1, $\overline{x2}$ is the mean of sample 2, $\sigma$ is the population standard deviation, while $n1$ and $n2$ are sizes of sample 1 and 2.

$F = \sigma_b^2 \div \sigma_w^2$, where $\sigma_b^2$ is the between column variance among sample means (treatment) and $\sigma_w^2$ is the within column variance (error).

$TukeyHSD = q * sqrt(MS_{within} \div n)$, where $MS_{within}$ is the mean square output from the ANOVA computed, $n$ is the total number of data points for a particular group, and $q$ is the studentized range statistic.

**4.1 Results**

Figure 6 depicts the variation of IPI, among high versus low engagers and in-video dropouts versus non dropouts, in the same video lecture 4-6 from the course, that we have been performing our experiments on. Similar findings were also confirmed with other randomly chosen course videos. Figure 7 shows the frequency distribution of IPI. These figures concur with our intuitions. The average IPI is significantly higher for students with "High" engagement ($|z|=8.296$, $p<0.01$) and "Non dropouts" ($|z|=22.54$, $p<0.01$). This is also reflected in the histogram, which clearly shows that many non dropouts have positive IPI that pushes up the average.

In order to generalize these findings, we also look at the variations of IPI among some other student partitions that we made for the whole course. "Viewers" are students who have watched or interacted with some video lecture but have not done the exercises; the "Active" students additionally turn in homework also. Among the "Active" participants, some students get the "Statement" of accomplishment. Students who achieve a grade (weighted sum of quizzes and assignments completed) of 80% or more achieve a "distinction"; students who achieve a grade from 60-80% are classified into "Normal" category; students having grade less than 60% qualify into "None" achievement category. MOOC dropouts are those students who cease to actively participate in the MOOC (we are concerned with video lecture viewing only) before the last week, i.e., students who do not finish the course.

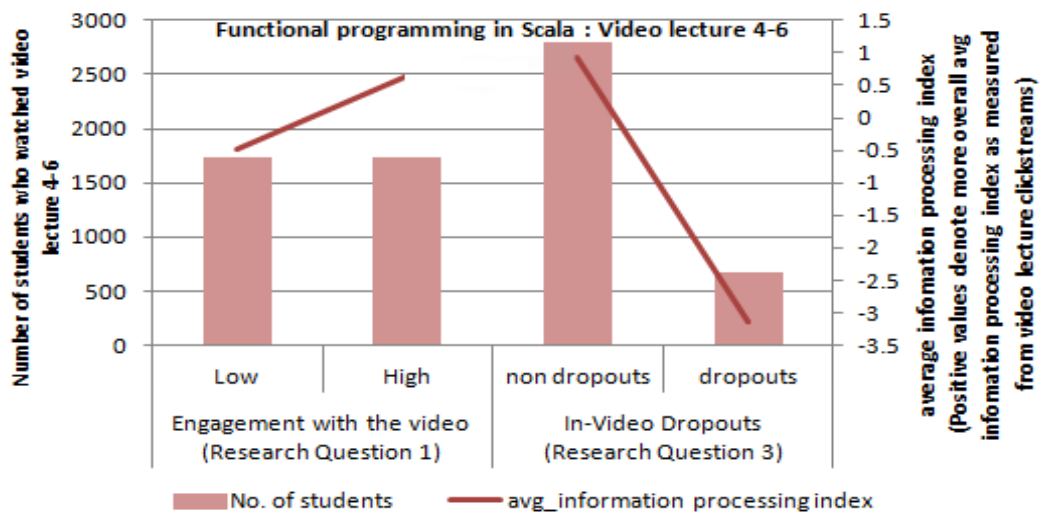

Fig 6: Variation of Average Information Processing Indices(IPI) for Video 4-6

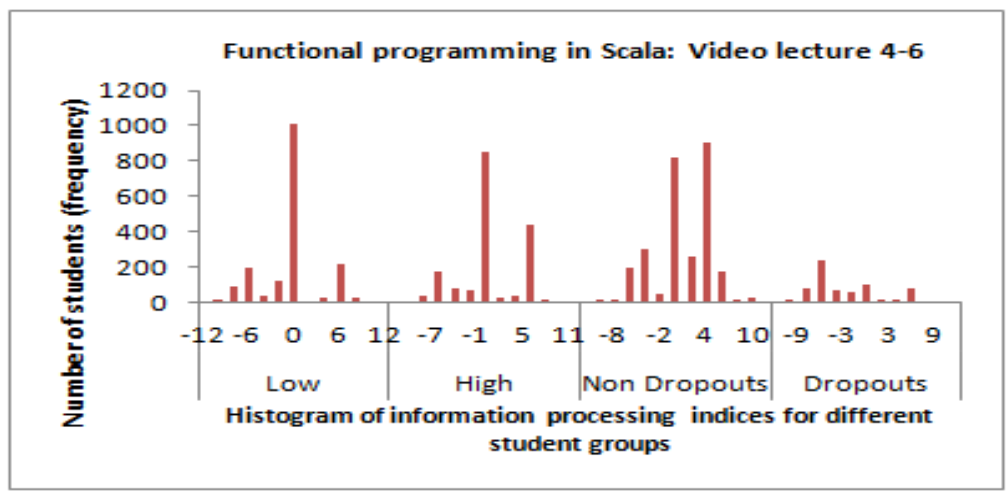

Fig 7: Frequency Distribution(histogram) of Information Processing Indices(IPI) for Video 4-6

An important observation in figure 8 is that IPI is clearly able to distinguish between Non-dropouts and Dropouts ($|z|=9.06$, $p<0.01$). This is also reflected in the histogram in figure 9, which verifies that more proportion of "Non dropouts" have positive IPI. More is the information processing done by students, greater is the video lecture involvement, higher are the chances to derive true utility from video lecture and remain excited and motivated to stay in the course. In addition, we also obtain striking differences between "Active" versus "Viewers" ($|z|=10.45$, $p<0.01$). Intuitively too, we expect "Viewers" to have higher IPI than "Active" class, because as their primary MOOC activity, "Viewers" grapple more with the video lecture.

In figure 8, we observe that students who do not achieve a statement have significantly higher IPI than the ones who get a statement ($|z|=4.58$, $p<0.01$). However, achieving a statement requires students to compulsorily complete course quizzes and assignments, in addition to watching MOOC video lectures (which of course is not compulsory and carries no credit). Therefore IPI alone is not a very good measure to distinguish "Statement" versus "No Statement" student groups. Similar argument holds for the three classes of "Achievement". Though the differences in mean values for "Distinction", "None" and "Normal" groups are significant ($F(2,19525)=11.16$, $p<0.0001$), we must be careful in interpretation, because the definitions for this partition are based fully on course grades (which of course will be partly affected by video lecture viewing).

A Tukey HSD posthoc test for "Achievement" bins reveals that pairwise differences in mean for "Distinction", "None" ($p<0.05$) and "Normal", "None" ($p<0.01$) are significant. The change in IPI across continents gives an insight into how significantly ($F(4,18215)=3.43$, $p=0.008$) this operationalization varies by demographics. For continents, a Tukey HSD posthoc test reveals that pairwise contrasts of Africas with Americas ($p<0.05$), Europe ($p<0.01$) and Oceania ($p<0.05$) are significant. All statistics computations are performed, with references from (Lowry, 1998).

While (DeBoer et al., 2013) have studied how diversity in MOOC students' demographics and behaviors is correlated to course performance and success, explicit background data on students was collected via an exit survey, rather than developing an implicit metric to measure performance.

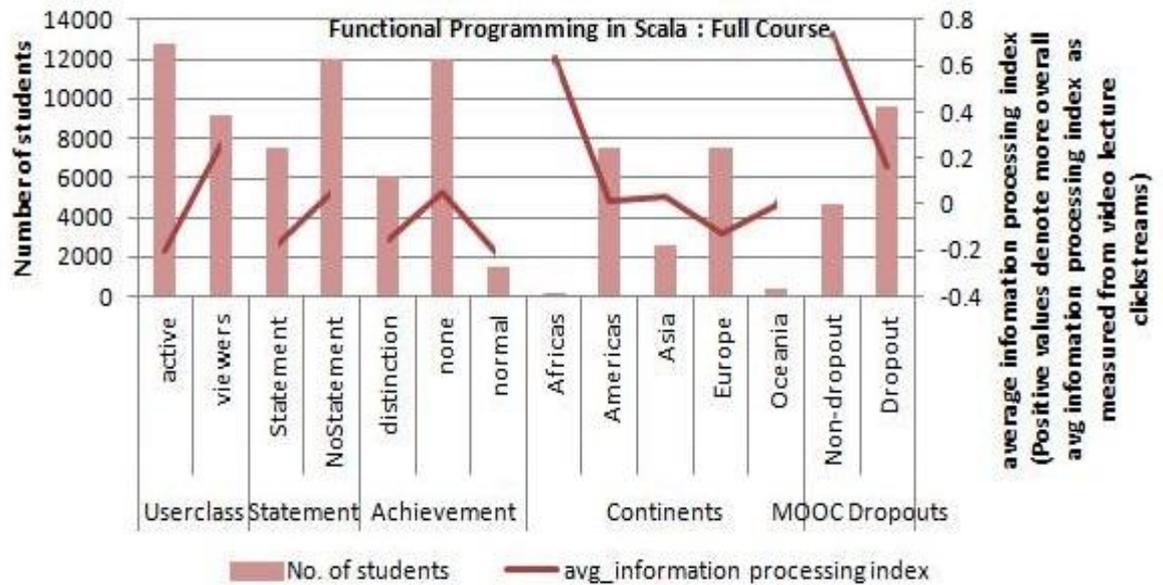

Fig 8: Variation of Average Information Processing Indices(IPI) for different student partitions in the full course

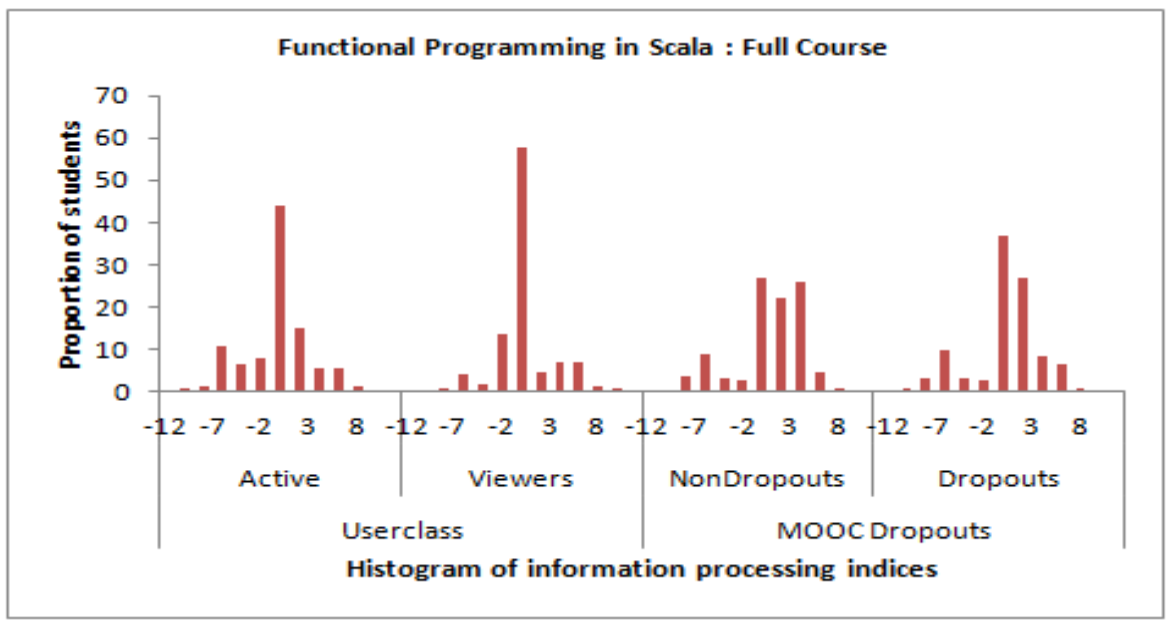

Fig 9: Frequency Distribution(histogram) of Information Processing Indices(IPI) for different student partitions in the full course

# CHAPTER 5

# VALIDATION EXPERIMENTS 3: STUDENT MODELING

## 5.1 Preliminaries for Approach 1

**A) Markov Chains:** Markov chains are stochastic process with states & transitions. A Markov chain of order 'm' (or a Markov chain with memory 'm'), where m is finite, is a process where the future state depends on the past 'm' states. This is diagrammatically represented in figure 10.

$$\Pr(X_n = x_n \mid X_{n-1} = x_{n-1}, X_{n-2} = x_{n-2}, \ldots, X_1 = x_1)$$
$$= \Pr(X_n = x_n \mid X_{n-1} = x_{n-1}, X_{n-2} = x_{n-2}, \ldots, X_{n-m} = x_{n-m}) \text{ for } n > m$$

Fig 10: Markov chain with a sequence of random variables $X_1, X_2, X_3\ldots$

Based on the sequence of events, a transition probability matrix is formed, and the next sequence of states can be predicted from the current state, using the formula: $x_{n+k}(t) = x_n(t).\mathbf{P}$, where 'P' is the transition matrix derived from fitting the Markov chain of a particular order. For example, the following figure 11 represents the $1^{st}$ order transition matrix (P) for some students' clickstream sequence.

|     | Pl | Pa | Sf | SSf | Sb | SSb | Rf | Rs |
|-----|----|----|----|----|----|----|----|----|
| Pl  | 0.000405 | 0.732392 | 0.05654 | 0.038926 | 0.183846 | 0.072963 | 0.004343 | 0.011015 |
| Pa  | 0.831106 | 0.1978 | 0.215556 | 0.160115 | 0.284444 | 0.239813 | 0.185464 | 0.223131 |
| Sf  | 0.059103 | 0.007333 | 0.531292 | 0.418408 | 0.102347 | 0.119493 | 0.015347 | 0.008812 |
| SSf | 0.006885 | 0.000817 | 0.070925 | 0.253883 | 0.013996 | 0.064427 | 0.001882 | 0.002832 |
| Sb  | 0.072445 | 0.057756 | 0.107261 | 0.088974 | 0.37539 | 0.343337 | 0.014333 | 0.031943 |
| SSb | 0.008099 | 0.003132 | 0.010603 | 0.033749 | 0.033445 | 0.153634 | 0.001593 | 0.003777 |
| Rf  | 0.013463 | 0.000202 | 0.005593 | 0.00326 | 0.003849 | 0.003029 | 0.477052 | 0.250197 |
| Rs  | 0.008495 | 0.000567 | 0.002231 | 0.002685 | 0.002683 | 0.003304 | 0.299986 | 0.468293 |

Fig 11: Transition Probability Matrix for a students' clickstream sequence, derived from a Markov chain of order 1

To measure the goodness of fit of a Markov chain, we use 3 criteria: A) Log Likelihood value, which in simple terms, measures the goodness of fit of a Markov chain (ideally, we want this value to be higher or closer to 0), B) Akaikes Information Criterion (AIC), which is a form of penalized log likelihood. AIC is an estimate of a constant plus the relative distance between the unknown true likelihood function of the data and the fitted likelihood function of the model, so that a lower AIC means a model is considered to be closer to the truth. AIC is better in situations when a false negative finding would be considered more misleading than a false positive, C) Bayesian Information Criterion (BIC), which is a slight variation of the penalized log likelihood. BIC is an estimate of a function of the posterior probability of a model being true, under a certain Bayesian setup, so that a lower BIC means that a model is considered to be more likely to be the true model. BIC is better in situations where a false positive is as misleading as, or more misleading than, a false negative.

The AIC or BIC for a model is usually written in the form [-2logL + kp], where L is the likelihood function, p is the number of parameters in the model, and k is 2 for AIC and log(n) for BIC.

**B)K Means Clustering:** Clustering is an unsupervised machine learning problem where the objective is to find hidden structure in unlabeled data. K means clustering is a popular centroid based clustering algorithm. Concretely, given a set of observations ($x_1$, $x_2$, ..., $x_n$), where each observation is a 'd-dimensional' real vector, k-means clustering aims to partition the 'n' observations into 'k' sets (k ≤ n), S = {$S_1$, $S_2$, ..., $S_k$}, so as to minimize the within-cluster sum of squares (WCSS). This is represented in figure 12. The algorithm is described in figure 13.

$$\arg\min_{\mathbf{S}} \sum_{i=1}^{k} \sum_{x_j \in S_i} \|x_j - \mu_i\|^2$$

Fig 12: Principle behind K Means clustering algorithm (minimizing within-cluster sum of squares). $\mu_i$ is the mean of points in $S_i$.

```
K-means algorithm

Randomly initialize K cluster centroids $\mu_1, \mu_2, \ldots, \mu_K \in \mathbb{R}^n$

Repeat {
    for $i$ = 1 to $m$
        $c^{(i)}$ := index (from 1 to K) of cluster centroid
                    closest to $x^{(i)}$
    for $k$ = 1 to K
        $\mu_k$ := average (mean) of points assigned to cluster $k$
}
```

Fig 13: The K Means clustering algorithm

**5.2 Approach 1: Markov Clustering of Student Clickstreams**

So far, we can intelligibly contemplate that IPI differs for different student partitions. As a matter of fact, this actually happens because of presence of smaller substructures inside these larger groupings, that are similar in their click behaviors. Deciphering these smaller clusters would be very meaningful for course instructors. It would aid in designing customized learning solutions for students within these clusters, who interact in unique ways with MOOC video lectures.

To demonstrate our approach, we randomly choose 6 videos from the course (1-4, 2-3, 2-5, 6-4, 7-5, 4-6). Firstly, we fit a Markov chain of order 1 to each students' clickstream (Log Likelihood: -303714.2, Akaikes Information Criterion (AIC): 607446.3, Bayesian Information Criterion (BIC) : 607543.5). This chain has the maximum log likelihood and minimum value of AIC, BIC (Dziak et al., 2012) when compared to Markov chains from order 2 to 5. The output is a transition probability matrix for each clickstream sequence. In the next step, we present these markov matrices as input to K-means clustering algorithm. The motivating intuition is to group similar matrices, having lot of click overlap (accounts for order and number) and similar transition probabilities. On varying 'k' from 4 to 9, k=8 gives minimum within cluster sum of squares and maximum between cluster sum of squares. The proportion of clicks belonging to each raw click category is presented in figure 14. Cluster attributes such as the time spent on seek forward, seek backward and pause are depicted in figure 15.

C1 and C2 represent normal watchers who primarily play and pause without doing much activities. However, the average clickstream sequence length for C1 is four times C2, and that is why these two clusters are differentiated. Cluster C3 represents watchers with low proportion of seek/scroll forward and seek/scroll backward clicks, while cluster C7 represents watchers with high proportion of seek/scroll forward and seek/scroll backward clicks. Average clickstream sequence length for C7 is 1.7 times that of C3. Next, clusters C4 and C6 represent watchers having high proportion of seek/scroll backward clicks, representative of revision or rewatching. Average clickstream sequence length for C6 is 1.8 times that of C4. On the contrary, cluster C5 represents watchers having high proportion of seek/scroll forward clicks, representative of skipping. Cluster C8 represents watchers who mainly do ratechange clicks. Such students are very likely, not to seriously follow the video lecture.

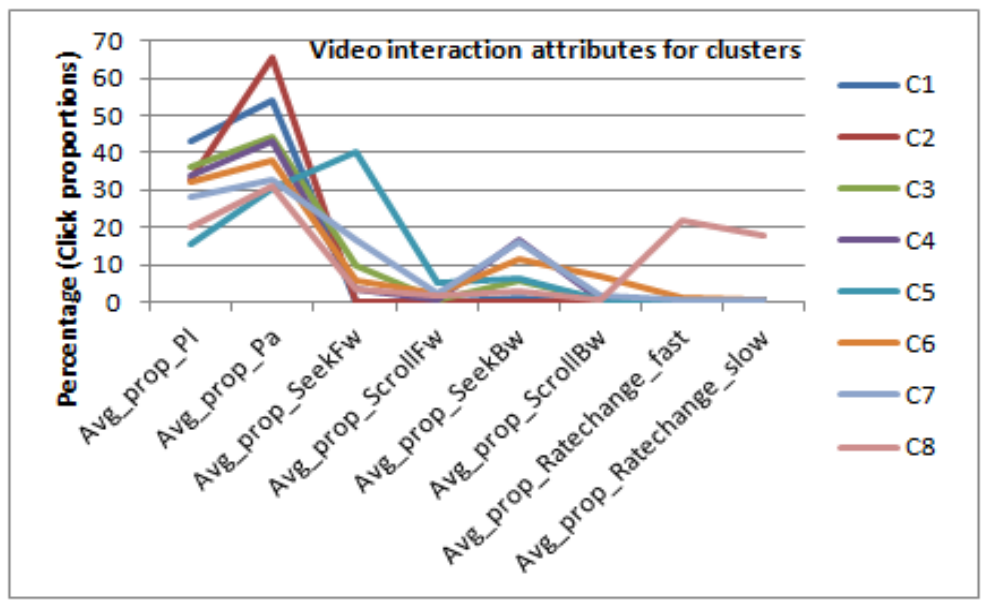

Fig 14: Differentiated clusters based on distinct video watching preferences of students in the MOOC

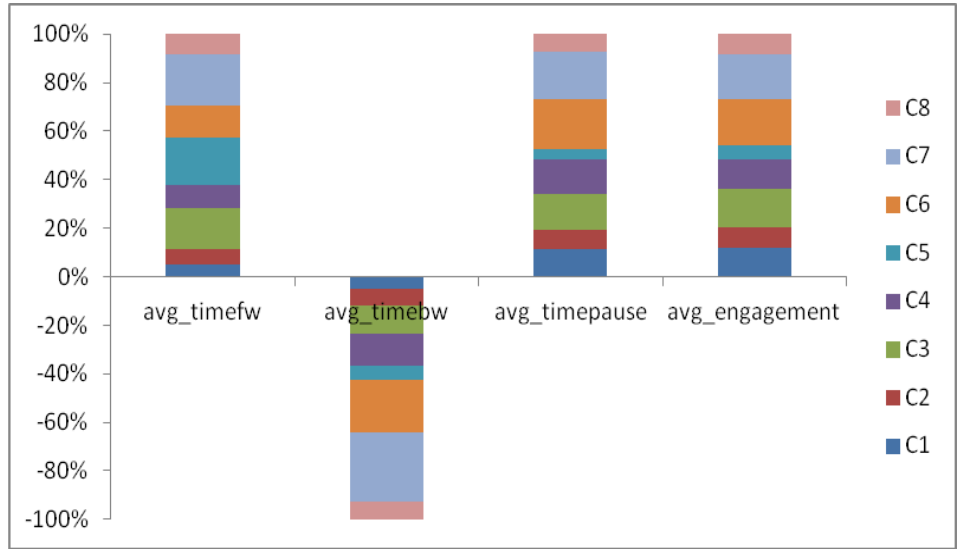

Fig 15: Cluster attributes such as the time spent on seek forward, seek backward and pause and total time spent on interacting with the video (engagement)

## 5.3 Preliminaries for Approach 2

Statistics provide useful tools for summarizing large amounts of social network information, and for treating observations as stochastic, rather than deterministic outcomes of social processes. When we use statistics to describe network data, we are describing properties of the distribution of relations or ties among actors, rather than properties of the distribution of attributes across actors.

Standard statistical tools for the analysis of variables cannot be directly applied to inferential questions, hypothesis or significance tests, because the individuals embedded in a network are not independent observations drawn at random from some large population. The "boot-strapping" approach (estimating the variation of estimates of the parameter of interest from large numbers of random sub-samples of actors) is used to get more correct estimates of the reliability and stability of estimates (i.e. standard errors).

Concretely, to examine relations between 2 types of ties in a network, essentially, we have two adjacency matrices, one for Type X ties and one for Type Y ties, and we would like to correlate them. We cannot do this using a standard statistical package for two reasons. First, statistical packages are set up to correlate vectors, not matrices. This is not a very serious problem, however, because we could just reshape the matrices so that all the values in each matrix were lined up in a single column with NxN values. We could then correlate the columns corresponding to each matrix. Second, the significance test in a standard statistical package makes a number of assumptions about the data which are violated by network data. For example, standard inferential tests assume that the data observations are statistically independent, which, in the case of matrices, they are not. To see this, consider that all the values along one row of an adjacency matrix pertain to a single node. If that node has a special quality, such as being very anti-social, it will affect all of their relations with others, introducing a lack of independence of all those cells in the matrix. Another typical assumption of classical tests is that variables are drawn from a population with a particular distribution, such as a normal distribution. Often times in network data, the distribution of the population variables is not normal or is simply unknown. Moreover, the data is probably not a random sample or even a sample at all ; all we have is a population

The QAP correlation/regression technique correlates two or more adjacency matrices by effectively reshaping them into two long columns and calculating an ordinary measure of statistical association such as Pearson's r. We call this the observed correlation. To calculate the significance of the observed correlation, the method compares the observed correlation to the correlations between thousands of pairs of matrices that are just like the data matrices, but are known to be independent of each other. To construct a p-value, it simply counts the proportion of these correlations among independent matrices that were as large as the observed correlation. As usual, we typically consider a p-value of less than 5%/1% to be significant (i.e., supporting the hypothesis that the two matrices are related).

QAP Regression allows us to model the values of a dependent variable (such as Type X ties) using multiple independent variables (such as Type Y ties and some other relations such as Type Z ties). The randomly generated pairs of adjacency matrices for each permutation are done by randomly rearranging rows and columns (therefore independent), rather than changing individual matrix entries. It has 2 advantages: Old and new matrices have same properties such as mean, standard deviation (s.d) etc. More subtle and auto-correlational properties of matrices are preserved, so when we compare the observed correlation against our distribution of correlations, we can be sure we are comparing apples with apples.

**5.4 Approach 2: Social Network Analysis based modeling**

In order to gain better visibility into how students in MOOCs are informally connected through a common pattern of clickstream interaction, we now present a social network analysis based student modeling. Specific questions that guide our work going forward include, a study on the significance of influencing relations (similarity in the proportion of video watched, engagement with the video, average playing rate or difficulty rating for the video) and video interaction attributes (number of seeks/pause, time spent on pause/seek) that affect the relationship between students having similar clickstream sequences. The data for this social network based student modeling comes from video lecture 4-6 (6th lecture in the 4th week of the course). Our steps to set up this analysis are as follows:

1) Firstly, we discretize various video interaction features:
- **Engagement attribute** = (summation of time spent on pause, seekFw and seekBw) * average play rate. This is discretized by equal frequency into 2 bins: 1(Low or $\leq$1112 secs), 2(High or >1112 secs)
- **Video played proportion attribute** = (played length/total video length) * average play rate * 100. This is discretized by equal width into 4 bins: 1(<51.105%), 2(51.105%, 100.737%), 3(100.737%, 150.369%), 4(>150.369%)
- **Average play rate attribute**. This is discretized by equal frequency into 2 bins: 1(Low or $\leq$ 1), 2(High or $\geq$1)

2) Then, we form 4 different kinds of network from the clickstream data for our experimentation purposes:

- **1st network (VWSS)**: 2 students connected if their video watching state sequences (VWSS) are similar/belong to same cluster (density: 0.45)
- **2nd network (VPP)**: 2 students connected if their video play proportion is similar/belongs to same cluster (density: 0.49)
- **3rd network (ET)**: 2 students connected if their engagement with the video is similar/belongs to same cluster (density: 0.49)
- **4th network (APR)**: 2 students connected if their avg playing rate is similar/belongs to same cluster (density: 0.56)

{To quantitatively define similarity of VWSS, we represent each VWSS using 8 numeric metrics such as proportion of Pl/Pa/Sf/Sb/Rc clicks, timeonPause/SeekFw/SeekBw. Then, K-Means clustering is applied to find similar VWSS (Distance metric: Euclidean, Scoring metric: Distance to Centroids). After optimization, we group VWSS into 4 clusters (k=4). So, two students will be connected, if their VWSS belong to the same cluster}

3) To motivate our **Dyadic Hypothesis**, firstly we combine the individual networks into multiplex relations, and examining the overall density and density within groups. We form the multiplex relation quantitatively using boolean combinations. For e.g: If there was a link between student A and student B because of having similar VWSS (1), AND there was also a link between student A and student B because of having similar ET (1), then the multiplex relation adjacency matrix would also have a 1 in the (ij)$^{th}$ entry corresponding to (student A, student B). All the 3 combinations in table 4 below were constructed similarly.

| Multiplex relation | Overall density | Density by groups | | | Hypothesis formed |
|---|---|---|---|---|---|
| **VWSS-VPP** (1st network AND 2nd network) | 0.27 | Number | Density | VPP | Students in Partition 1 and 2 have max no. of ties with high density. They should have similar VWSS. |
| | | 1  25874.000 | 0.631 | <51.105% | |
| | | 2  3029354.000 | 0.573 | 51.105%,100.737% | |
| | | 3  200862.000 | 0.322 | 100.737%,150.369% | |
| | | 4  9872.000 | 0.272 | >150.369% | |
| **VWSS-ET** (1st network AND 3rd network) | 0.22 | Number | Density | ET | Both partitions have almost equal density. Overall density is also low. Therefore, these students need not have similar VWSS. |
| | | 1 1420140.000 | 0.469 | Low | |
| | | 2 1301148.000 | 0.429 | High | |
| **VWSS-APR** (1st network AND 4th network) | 0.29 | Number | Density | APR | Partition 1 has much greater no. of ties with very high density as compared to Partition 2. Therefore, students in partition 1 should have similar VWSS. |
| | | 1 3153428.000 | 0.573 | <=1 | |
| | | 2 369764.000 | 0.287 | >1 | |

Table 4: Examining density in multiplex network relations to motivate dyadic hypothesis

An index based on multiplex relations is therefore of considerable interest, because it tells us about which groups of students are prevalent and which subgraphs do we select for closer analysis. Now, we test whether the above hypothesis formed are significant.

Our Dyadic Hypothesis (the more a pair of persons has a certain kind of relationship, what is the increase/decrease in likelihood that they will also have another kind relationship) is: Are there relationships associated with, which students have similar video watching sequences? We imagine that students cannot have similar video watching sequences to others randomly. Influencing factors might be having similarity in the proportion of video watched, engagement with the video in secs and avg playing rate for the video. Do the pattern of ties among these multiple relations align?

The results summarized in Figure 16, conform with the motivation presented in Table 4. The model R square indicates that knowing whether 2 students have similar video play prop/engagement/average playing rate reduces uncertainty in predicting whether they will also have similar VWSS by about 4%. Though the factors are not determining factors, still, coefficient for 2nd/4th network are significant in the output (positive relationship at 0.01 LOS): indicating that students who have similar VWSS, have similar video play prop and similar avg playing rate.

Since the dependent matrix is binary, the regression equation can be represented as a linear probability model. The intercept tells us that if 2 students do not have same video play prop/engagement/average playing rate, the probability of 2 students having similar VWSS is 0.34. If the students have same video play proportion, this increases the probability of similar VWSS by 0.160. If the students have same average playing rate, this increases the probability of similar VWSS by 0.045. For example: In a batch of say, 1000 student dyads where i and j have similar video play prop/avg playing rate, we expect to see about 160/45 more cases of similar video watching sequences than when i and j don't have similar video play prop/avg playing rate.

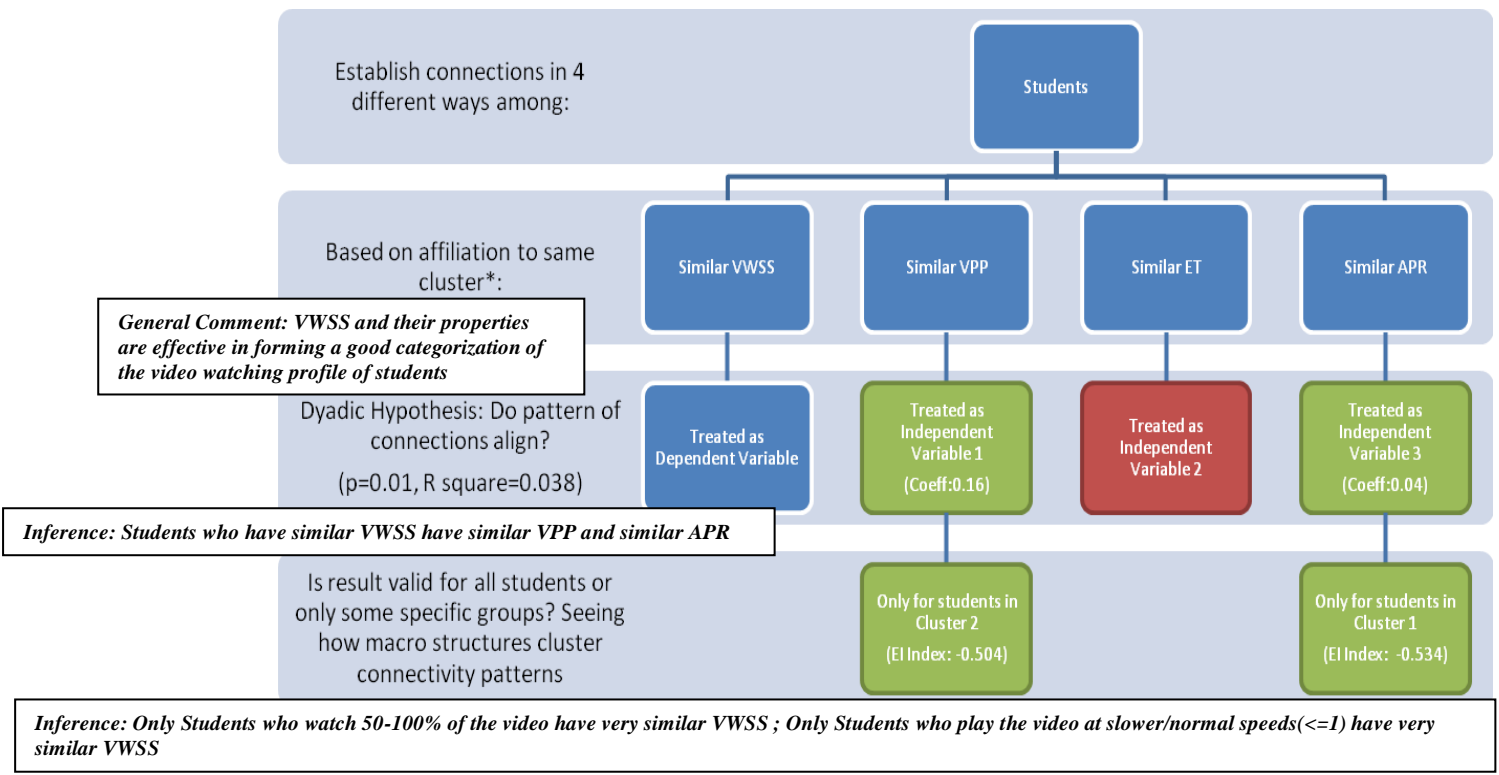

Fig 16: Hierarchical summary of validation results for dyadic social network hypothesis

To understand Level 4 of figure 16, we must understand that students may be embedded in macro structures while watching MOOC video lectures, and some of these macro structures can be the interaction attributes that characterize different aspects of video watching behavior. The E-I index/homophily measures quantify the extent to which these macro-structures cluster the connectivity patterns of individuals who fall within them, by telling what similarities/differences exist between students in their video watching behavior at a finer level of granularity.

Concretely, given a partition of a network into a number of mutually exclusive groups, E-I index/Homophily is defined as the number of ties external to the groups minus the number of ties that are internal to the group divided by the total number of ties. This value can range from 1 to -1 and can be seen as a measure of the extent a group chooses themselves. A value of -1 shows homophily and a value of +1 showing heterophily. For valued data it is the sum of the tie strengths instead of the number of ties.

Therefore, when we partition the 1st network based on attributes of 2nd, 3rd and 4th network, we see whether all sub-partitions are homphilous, or the pattern of connections align for only some of the smaller macro structures embedded.

4)To set up the next experiment, we form certain additional networks based on video interaction attributes of students in MOOCs. To motivate our **Monoadic Hypothesis**, firstly we combine the individual networks constructed from video interaction attributes (number of pauses/seekFw/seekBw) into multiplex relations, examining the overall density, and the density within groups. We form the multiplex relation quantitatively using boolean combinations. For e.g: If there was a link between student A and student B because of having similar VWSS (1), AND there was also a link between student A and student B because of having same number of Pauses(1), then the multiplex relation adjacency matrix would also have a 1 in the (ij)th entry corresponding to (student A, student B). All the 3 combinations in the table 5 below were constructed similarly.

| Multiplex Relation | Overall density | Hypothesis formed |
|---|---|---|
| **1st network AND SamePauses** | 0.11 | |
| **1st network AND SameFWs** | 0.31 | Should be significant |
| **1st network AND SameBWs** | 0.28 | Should be significant |
| **2nd network AND SamePauses** | 0.11 | |
| **2nd network AND SameFWs** | 0.24 | |
| **2nd network AND SameBWs** | 0.23 | |
| **3rd network AND SamePauses** | 0.15 | |
| **3rd network AND SameFWs** | 0.23 | |
| **3rd network AND SameBWs** | 0.25 | |
| **4th network AND SamePauses** | 0.12 | |
| **4th network AND SameFWs** | 0.26 | Should be significant |
| **4th network AND SameBWs** | 0.27 | Should be significant |

Table 5: Examining density in multiplex network relations to motivate monoadic hypothesis

Our monoadic hypothesis is: What attributes (number of seekFw/seekBw/Pauses, time spent on Pause/seekFw/seekBw) might predict which students have same video play prop/engagement/average play rate? In other words, what is more predictive of having same video play prop/engagement/average play rate: having similar number of seekFw/seekBw/Pauses, or the time spent on Pause/seekFw/seekBw?

Examining such a hypothesis focuses directly on a very fundamental sociological question: What factors affect the likelihood that 2 students will have a relationship? There are 2 social theories that link dyadic data with monoadic attributes:

A) **Homophily hypothesis**: There might be more ties between students having same video interaction attributes than we would expect by chance. We hypothesize that students have a tendency to be connected to other students having similar video interaction attributes as themselves, a phenomenon known as Homophily. Homophily is an instance of a larger class of frequently hypothesized social processes known as selection, in which actors choose other actors based on attributes of those actors.

B) **Diffusion hypothesis**: Diffusion is the idea that people's beliefs, attitudes, practices and so on, come about in part because of interaction (virtue of being connected) with others

The difference between diffusion and selection hypotheses is just the direction of causality. In diffusion, the dyadic variable causes the monadic variable, and in the selection the monadic variable causes the dyadic variable. So, in our case, for the MOOC data, our experimental design is motivated by the selection hypothesis. The standard approach to testing the association between a node attribute and a dyadic relation is to convert the problem into a purely dyadic hypothesis by constructing a dyadic variable from the node attribute. Different techniques are needed depending whether the attribute is categorical, such as gender or department, or continuous, such as age or wealth, which locate nodes along a continuum of values. In our case, for the MOOC data, we calculate "Exact matches" in video interaction attributes for each pair of students. A value of 1 indicates that 2 students have exactly same attributes, and value of 0 indicates different attributes. The results are summarized in figure 17.

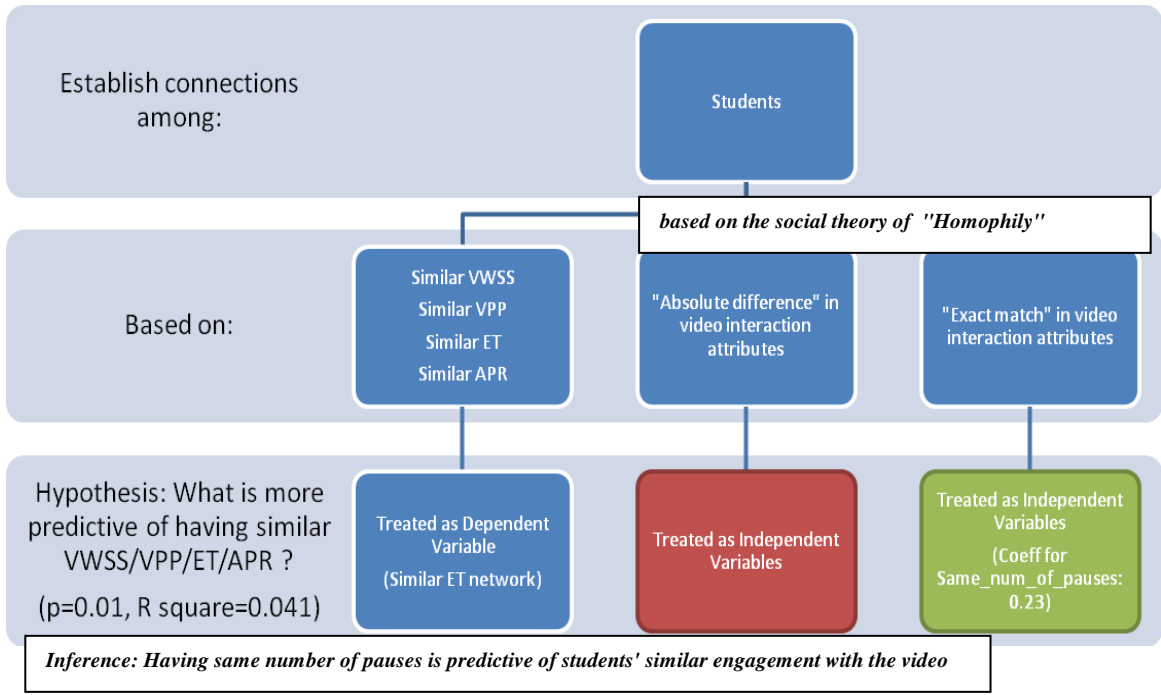

Fig 17: Hierarchical summary of validation results for monoadic social network hypothesis

# CHAPTER 6

# PREDICTING STUDENT ATTRITION USING VIDEO WATCHING BEHAVIOR

**6.1 Variation of dropouts in the MOOC**

We may expect that when students find the course too tough to follow/uninteresting/boring, they will not engage with future videos, or, when students seem very interested in understanding the video and exhibit lot of rewatching behavior, we might expect them to stay on through the course end video lecture. Therefore, students who do not stay till the last week of the course (exhibit any video lecture viewing), are considered as complete course dropouts.

We plot figures 18 and 19 to depict the proportion of course dropouts by weeks and by different student groups as discussed in previous chapters. Figure 18 concurs with our intuitions. Student groups "Active", "Statement" and "Distinction" have significantly lower dropout proportion (number) than students who lie in the respective complementary bins: "Viewers" ($\chi2$=1864.897, df=1, p<0.0001) , "No Statement" ($\chi2$=2105.066, df=1, p<0.0001), "Normal/None" ($\chi2$=5746.8, df=2, p<0.0001). The $\chi2$ value indicates that number of dropouts/non dropouts vary significantly in these distinct student partitions of the MOOC. In a $\chi2$ test, standardized residuals are a kind of z-score indicating how many standard deviations above or below the expected count a particular observed count is. For our experiment, the standardized residuals (z) for "all" categories in the contingency table which depicts the number of dropouts/non dropouts among A)Active versus Viewers, B)Statement versus No Statement, C)Distinction versus Normal versus None, are significant at 1% level of significance [This is equivalent to testing the null hypothesis that the actual frequency equals the expected frequency for a specific cell versus the research hypothesis of an absolute difference greater than zero]

$$z = \frac{|\text{observed} - \text{expected}| - 0.5}{\sqrt{\text{expected}}}$$

Moreover, the proportion (number) of dropouts slightly varies among continents ($\chi2$=87.58, df=4, p<0.0001), as indicated by the low $\chi2$ value. However, "not all" categories in the contingency table which depicts the number of dropouts/non dropouts are significant. At 1% level of significance, cells in the contingency table which depicts the number of dropouts/non dropouts in Americas and Europe regions are significant. At 5% level of significance, cells in the contingency table which depicts the number of dropouts/non dropouts in Asia, Americas and Europe regions are significant.

If we analyze figure 18 together with figure 8 (variation of average information processing indices), we can observe that average IPI for "Viewers" is much higher than "Active " class, despite more dropout proportion. This indicates that though "Viewers" put higher effort and more cognitive processing to follow the video lectures, there is insufficiency in understanding the course instruction, as well as getting in sync with the instruction delivery method and its pace. Similar correspondence can be seen between the average IPI for "No Statement" class and their dropout proportion, as compared to "Statement" class of students.

The peaks (local maximas) in figure 19 highlight video lectures that are "not easy to follow" or are "unable to hold students' attention", because we lose comparatively higher students after these lectures. In this figure, we also notice that very high number of students drop out after the 1st week (introductory set of lectures). One possible explanation for this might be that such students register for the course, to just see what is the course is about, without having any actual intention to follow the course. Information about such students is very helpful for a course instructor to design motivating interventions to help them to follow the course. One principal utility of detecting dropouts early is recommendation of selected future video lectures for students to watch (for example, where an interesting concept/case study/application is going to be discussed)

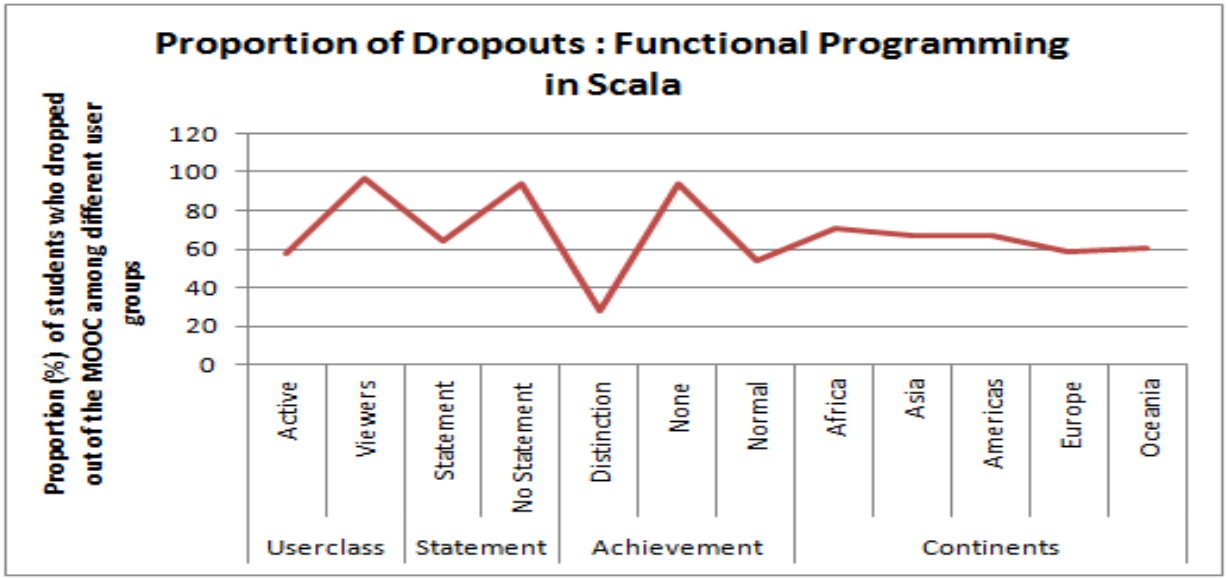

Fig 18: Proportion of dropouts among different student partitions in the MOOC

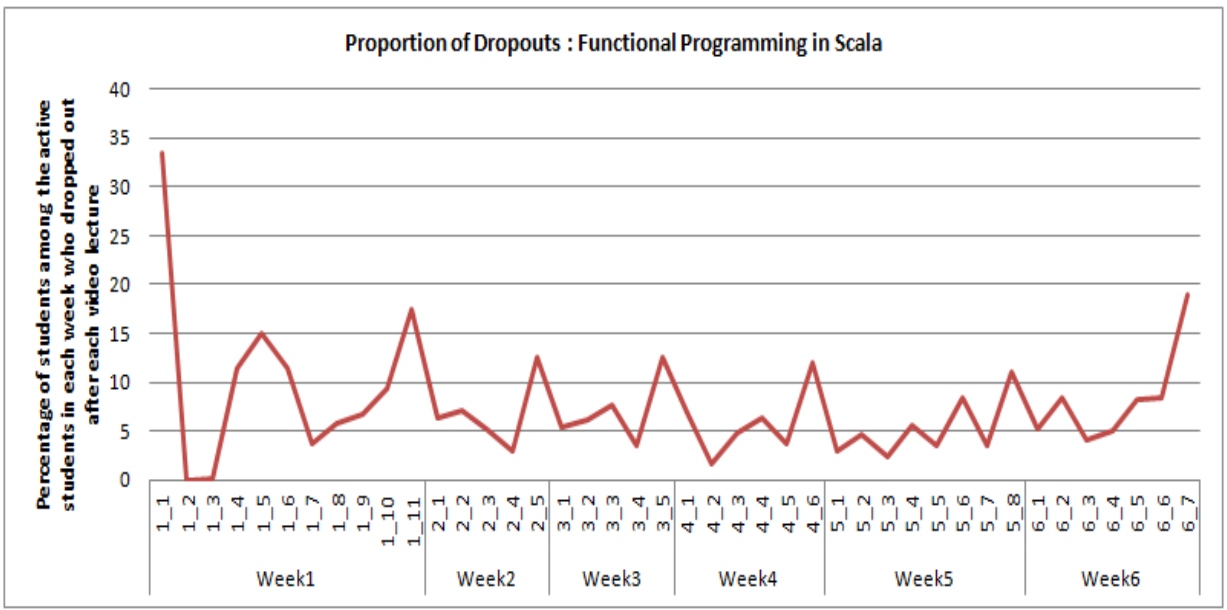

Fig 19: Proportion of dropouts by week (after each video lecture)

**6.2 Preliminaries for Dropout prediction**

Survival analysis (Miller, 2011) is a statistical modeling technique used to model the effect of one or more indicator variables at a time point on the probability of an event occurring on the next time point. In our case, we are modeling the effect of certain video interaction attributes (such as summarized clickstream behavior, information processing index, average playing rate etc) on probability that a student drops out of the video lecture participation on the next time point. Survival models are a form of proportional odds logistic regression, and they are known to provide less biased estimates than simpler techniques (e.g., standard least squares linear regression) that do not take into account the potentially truncated nature of time-to-event data (e.g., students who had not yet ceased their participation at the time of the analysis but might at some point subsequently). In a survival model, a prediction about the probability of an event occurring is made at each time point based on the presence of some set of predictors. The estimated weights on the predictors are referred to as hazard ratios. The hazard ratio of a predictor indicates how the relative likelihood of the failure (in our case, student dropout) occurring increases or decreases with an increase or decrease in the associated predictor.

A hazard ratio of 1 means the factor has no effect. If the hazard ratio is a fraction, then the factor decreases the probability of the event. For example, if the hazard ratio was a number n of value .4, it would mean that for every standard deviation greater than average the predictor variable is, the event is 60% less likely to occur (i.e., 1 - n). If the hazard ratio is instead greater than 1, that would mean that the factor has a positive effect on the probability of the event. In particular, if the hazard ratio is 1.25, then for every standard deviation greater than average the predictor variable is, the event is 25% more likely to occur (i.e., n -1).

**6.3 Dropout prediction**

    **6.3.1 Machine Learning Approach**

Having seen how dropout proportion across weeks and across different student partitions in the MOOC, we now seek to understand more about how participation trajectories of complete course dropouts differs from non dropouts. Therefore, it is interesting to investigate, the extent to which engagement, video play proportion and IPI trajectories influence attrition behavior. The development of trajectories is indicated in Table 6. Engagement (time in seconds) of a student is discretized by equal frequency into High and Low categories, considering all interactions each video lecture in the MOOC separately (because length of each video differs, so the discretization criteria would also differ for each video). Video play

proportion((video played length/video length)*100*average play rate) for a student is discretized by equal width (Very Low: <50%, Low: 50-100%, High: 100-150%, Very High: >150%). IPI for a student is discretized by equal frequency (Very Low: <-1.00, Low: [-1.00, 1.00], High: [1.00, 3.00], Very High: >3.00).

| Videos Watched | Video 1 | Video 5 | Video 6 | Video 9... |
|---|---|---|---|---|
| **Engagement** | High | Low | Low | High.. |
| **Engagement Trajectory** | H L L H... | | | |
| **Video Play Prop (Vpp)** | High | Very Low | Low | Very High.. |
| **Engagement Trajectory** | H VL L VH... | | | |
| **IPI** | Very High | Low | Very Low | High.. |
| **IPI Trajectory** | VH L VL H... | | | |

Table 6: Example depicting how different kinds of operationalized trajectories of students are formed

**Research Question:** Can we find some patterns in the video watching profile/behavior/trajectory of students, which can effectively say when are students most likely not to view the future video lectures?

**Settings:** The data for this experiment comes from 40 Videos of "Functional Programming in Scala" MOOC (4710 non-dropouts, 9596 dropouts). To address the skewed class distribution, cost sensitive L2 regularized Logistic Regression is used as the training algorithm (with 5 fold cross validation annotated by student-id and rare feature extraction threshold being 5). The dependent variable is the binary variable, complete course dropout (0/1). Dropout variable is 1 on the students' last week of active participation, and is 0 for all other weeks. If a students' final participation week is the last course week, dropout variable will remain 0 for that student for all weeks (the student is a non-dropout). To extract the interaction footprint of a student before he drops out of the course, we extract the following features: A)Transition features from "Engagement trajectory", "Video Play Proportion trajectory" and "IPI trajectories" of students for the videos watched (N-grams of length 4,5 and string length) from 0 to (n-1)$^{th}$ instant, B)Engagement, Video Play Proportion and IPI trajectories for the n$^{th}$ instance (attribute for the last video lecture watched before dropping out), C)Proportion of different symbol representations in the trajectories (for example, in a trajectory such as HLLHH, proportion(H)=60%, proportion(L)=40%.

**Results:** We achieve an accuracy of **0.80** and a kappa of **0.57** (Random baseline performance is 0.5). The false negative rate is **0.143**.

### 6.3.2 Survival Analysis

Using the statistical programming language R, we perform Survival analysis on our MOOC dataset. The variables we use are our quantitative IPI index, discretized engagement (high/low), discretized videoplayprop (low/medium/high/very high), jumped length forward (in secs), jumped length backward (in secs), summarized and discretized clickstream action vectors (rewatch, skipping, playratetransition, clearconcept, fastwatching, slowwatching, checkbackreference) and actual engagement (in secs). As an input, we standardize all the numeric variables (by computing z-scores). We transform the representation for "low" and "high" engagement to binary variable 0 and 1 to provide as an input to the survival model. Also, we transform "low", "medium", "high" and "very high" video play proportion categories into 0, 1, 2, 3. We remove all correlated variables, keeping only variables having less than 0.5 correlation for our analysis, to prevent multicollinearity problems.

| **Independent Variable** | **Hazard Ratios** |
|---|---|
| **IPI (Information Processing Index)** | 0.6367*** |
| **Rewatch behavioral action** | 0.6734*** |
| **Playrate Transition behavioral action** | 1.3585*** |
| **Video play proportion** | 0.6334*** |

Table 7: Hazard ratios of video interaction variables in the survival analysis (***: p<0.001)

The results are summarized in Table 7. Effects are reported in terms of the hazard ratio (HR), which is the effect of an explanatory variable on the risk or probability of participants drop out from the course, based on video lecture participation. Because all the explanatory variables except engagement/video play proportion have been standardized, the hazard rate here is the predicted change in the probability of dropout from the course forum for a unit increase in the predictor variable (i.e., Engagement changing from 0 to 1, or, Video play proportion changing by 1 unit (for example, from 0 to 1, 2, 3) or, the continuous variable increasing by a standard deviation when all the other variables are at their mean levels).

The hazard ratio for IPI means that students' dropout in the MOOC is 37% (100%-(100%*0.63)) less likely, if they have one standard deviation greater IPI than average. Such students grapple more with the course material (as reflected by their video lecture participation). Because video played proportion is a categorical variable, its hazard ratio tells us that increasing the video play proportion by 1 unit decreases the likelihood of student dropout by 37% (100%-(100%*0.63)). As students start watching more proportion of the video, this is indicative of their interest; as a result, they are less likely to dropout of the MOOC. Among other interesting results are the hazard ratios for rewatch and playrate-transition behavioral action. If students' rewatching behavior changes by 1 unit (from low to high), they are 33% (100%-(100%*0.67)) less likely to dropout. If students' playrate-transition behavior changes by 1 unit (from low to high), they are 35% ((100%*1.35)-100%) more likely to dropout. This indicates that such students have severe problems in coping up with the instruction pace and there is a definite lack of coherency between instruction pace and understanding.

In contrast to regular courses where students engage with class materials in a structured and monitored way, and instructors directly observe student behavior and provide feedback, in MOOCs, it is important to target the limited instructor's attention to students who need it most (Ramesh et al., 2013). By identifying students who are likely to end up not completing the class before it is too late, we can perform targeted interventions (e.g., sending encouraging emails, posting reminders, allocating limited tutoring resources, etc.) to try to improve the engagement of these students. For example, our prediction model could be used to improve targeting of limited instructor's attention to users who are motivated in general but are experiencing a temporary lack of motivation that might threaten their continued participation, in particular, those who have shown serious intention of finishing the course by interacting with a couple of video lectures.

# CHAPTER 7
# CONCLUSION AND FUTURE WORK

In this thesis work, we have begun to lay a foundation for research investigating students' information processing behavior while interacting with MOOC video lectures. The cognitive video watching model that we applied to develop a simple, yet potent IPI using linear weight assignments, can be effectively used as an operationalization for making predictions regarding critical learner behavior. As a next step, we plan on constructing a gradient function that captures the information processing hierarchy in a more robust manner. An additional challenge is to fuse video clickstreams with page-view clickstream gathered from the MOOC, to better understand students' interests during their interaction.

In our work going forward, we seek to understand how perceived difficulty of students (gathered in the form of a rating via an explicit questionnaire) is reflective of their engagement in the video and how it relates to high and low overall MOOC performance. Highlighting video lectures that are "not easy to follow" or are "unable to hold students' attention", would be helpful for a course instructor to design motivating interventions for students to follow the course. Another interesting enhancement to our work will include comparative analysis of the currently studied introductory level MOOC course, with intermediate and advanced level courses to contrast and generalize the findings.

We will draw from work integrating statistical approaches such as survival models and social network analysis techniques, in order to form combined representations of video lecture and page-view clickstream behavior as well as discussion forum footprint. This will help us to gain better visibility into how students participate in these MOOCs as a whole. Combining such students inputs with more granular behaviors such as eye tracking (Schneider et al., 2013) would help us to investigate deeply, the factors that influence students' interaction.

**APPENDIX**

Results of this thesis work (except insights from chapter 5) has been submitted to **"Modeling Large Scale Social Interaction in Massively Open Online Courses Workshop (Conference: Empirical Methods in Natural language Processing - EMNLP 2014)"**.

Concepts that have been discussed in chapter 5 and 6, are being currently applied to develop computational models (combining video clickstream and discussion forum activity behavior) for predicting student dropout in MOOCs in the "shared task" competition for the **"Modeling Large Scale Social Interaction in Massively Open Online Courses Workshop (Conference: Empirical Methods in Natural language Processing - EMNLP 2014)"**